\documentclass[12pt]{article}
\usepackage{graphicx}
\usepackage{epstopdf}

\newcommand{\jpsi}{J/$\psi$\hspace{0.1cm}}
\newcommand{\upsi}{$\Upsilon$\hspace{0.1cm}}

\newcommand{\pt}{$p_\mathrm{T}$\hspace{0.1cm}}
\newcommand{\kt}{$k_\mathrm{T}$\hspace{0.1cm}}
\newcommand{\bb}{$b\bar{b}$ }
\newcommand{\BB}{$B\bar{B}$ }

\newcommand{\pp}{$p\bar{p}$ }

\def\Title#1{\begin{center} {\Large #1 } \end{center}}
\def\Author#1{\begin{center}{ \sc #1} \end{center}}
\def\Address#1{\begin{center}{ \it #1} \end{center}}

\newenvironment{Abstract}{\begin{center}{\bf Abstract}\end{center} \bigskip \begin{quotation}}{\end{quotation}}
\newenvironment{Presented}{\begin{quotation} \begin{center} 
             PRESENTED AT\end{center}\bigskip 
      \begin{center}\begin{large}}{\end{large}\end{center} \end{quotation}}
\def\Acknowledgements{\bigskip  \bigskip \begin{center} \begin{large}
             \bf ACKNOWLEDGEMENTS \end{large}\end{center}}

\begin{document}
\begin{titlepage}


\Title{Heavy flavor physics with CMS}
\vfill
\Author{P. Bellan\\
on behalf of\\
the CMS Collaboration}  
\Address{Padova University and INFN -- 35131 Padova, Italy}
\vfill


\begin{Abstract}
\noindent
Recent results from CMS Collaboration on quarkonia physics and 
heavy quark production are presented. 
All these results have been obtained analyzing the data from $pp$ collisions at $\sqrt{s}=7$~TeV provided by the LHC and collected by the CMS detector in the year 2010.
The measurements of B-mesons, charmed meson and open beauty production 
cross sections are illustrated, together with the analysis techniques and
the estimation of the systematics uncertainties, and  
compared with the predictions of the available theoretical models. 
A recent result from CDF on the \upsi polarization is also reported. 
\end{Abstract}

\vfill

\begin{Presented}
The Ninth International Conference on\\
Flavor Physics and CP Violation\\
(FPCP 2011)\\
Maale Hachamisha, Israel,  May 23--27, 2011
\end{Presented}
\vfill

\end{titlepage}
\setcounter{footnote}{0}


\section{Introduction}\label{ref:intro}

\noindent
The study of heavy-quark production in high-energy hadronic interactions
plays a key role in testing next-to-leading order (NLO) 
Quantum Chromodynamics (QCD) calculations. 
In the past, discrepancies were observed between experimental data and theoretical predictions, e.g. at Tevatron~\cite{d0:inclmub, d0:bb, cdf:bsemilept, cdf:b2mud0x} 
and HERA~\cite{h1:openb, h1:muj, zeus:mudij, zeus:bc2mu}. 
Substantial progress has been achieved in the understanding of heavy-quark 
production at Tevatron energies~\cite{Cacciari04}, but large theoretical 
uncertainties still remain, mainly due to the dependence of the calculations on the renormalization and factorization scales.
The observed large scale dependence of the NLO calculations is considered
to be a symptom of large contributions from higher orders: small-$x$
effects~\cite{Collins:1991ty, Catani:1990eg}, where 
$x \sim m_\mathrm{b}/\sqrt{s}$, are possibly 
relevant in the low transverse
momentum (\pt) domain, while multiple-gluon 
radiation leads to large logarithms of \pt$/m_\mathrm{b}$ and may be
important at high \pt\cite{Cacciari:1993mq}. The resummed logarithms of
\pt$/m_\mathrm{b}$ at next-to-leading-logarithmic accuracy have been matched to
the fixed-order NLO calculation for massive
quarks~\cite{Cacciari:1998it}.  At the non-perturbative level, the b-hadron
\pt spectrum depends strongly on the parametrization of the
fragmentation function~\cite{Frixione:1997ma}.  The b-quark
production cross section has also been studied in the general-mass
variable-flavor-number scheme~\cite{Kniehl:2008zza} and the $k_\mathrm{T}$
factorization QCD approach~\cite{Ryskin:2000bz,Jung:2001rp}.  

Measurements of B-hadron production 
at higher energies than before, provided
by the Large Hadron Collider (LHC), represent an important 
test of the new theoretical calculations just mentioned. 
Measurements of inclusive b-quark production cross section 
require identification of inclusive events in which a 
b-quark has been produced in $pp$ collisions.
In results reported here, the discrimination of the heavy quark events 
has been achieved either exclusively, reconstructing the whole decay 
channel of a B meson, or
inclusively, considering hard jets. In this context, two different tagging techniques
have been applied: reconstruction of a displaced secondary vertex, and 
analysis of the transverse momentum spectrum of
an energetic muon with respect to the closest jet.  

Concerning the charmed and beauty bound states, it is well known that 
the mechanisms of quarkonia production
are still not fully understood, and 
despite of the considerable progress made in recent years, none of the
existing theoretical models describes satisfactorily 
prompt \jpsi and \upsi differential cross 
sections~\cite{yellow}, nor the
polarization values obtained with the Tevatron data. 
Therefore, measurements at the LHC 
will contribute to shed light to the quarkonium production mechanisms, by providing
differential cross sections in wider rapidity ranges   
and to higher transverse momenta than before. 
They also allow for important tests of several alternative theoretical 
approaches:
these include non-relativistic QCD (NRQCD) factorization~\cite{bib-nrqcd}, 
where quarkonium production includes colour-octet (CO) components, 
and calculations made in the colour-singlet (CS) model including 
next-to-leading order (NLO) 
corrections~\cite{Artoisenet:2008fc} which reproduce the 
differential cross sections 
measured at the Tevatron experiments~\cite{d0:upsipol, cdf:upsipol} 
without requiring a significant colour-octet contribution. \\
\noindent



\section{The CMS detector}
A detailed description of the CMS detector can be found elsewhere~\cite{JINST}.
Some of the most relevant features for heavy flavor physics are summarized here. 

The core CMS apparatus is a superconducting solenoid, of 6~m internal diameter, 
providing a magnetic field of 3.8~T.  
Within the field volume there are the silicon pixel and strip tracker, the crystal electromagnetic calorimeter and the brass/scintillator hadron calorimeter. 
Muons are detected by three types of gas-ionization detectors embedded in the steel return yoke: Drift Tubes (DT), Cathode Strip Chambers (CSC), and Resistive Plate Chambers (RPC).
The muon detectors cover a pseudorapidity window $|\eta|< 2.4$, where $\eta = - \ln [\tan (\theta / 2)]$, where the polar angle $\theta$ is measured from the $z$-axis, which points along the counterclockwise beam direction. 
The silicon tracker is composed of pixel detectors (three barrel
layers and two forward disks on each side of the detector, made of
66~million $100\times150$~$\mu$m$^2$ pixels) followed by microstrip detectors
(ten barrel layers plus three inner disks and nine forward disks on each side of the
detector, with 10~million strips of pitch between 80 and 184~$\mu$m).
Thanks to the strong magnetic field and the high granularity of the silicon tracker, the
transverse momentum, $p_\mathrm{T}$, of the muons matched to reconstructed
tracks is measured with a resolution of about 1\,\%  for the typical muons used in this analysis.
The silicon tracker also provides the primary vertex position, with $\sim$\,20~$\mu$m accuracy.  

The first level (L1) of the CMS trigger system, composed of custom
hardware processors, uses information from the calorimeters and muon
detectors to select the most interesting events.  The High Level
Trigger (HLT) further decreases the rate before data storage.

\section{CMS results on \jpsi and \upsi}
In the first weeks of the LHC operation at $7$ TeV in 2010, 
the CMS detector was able to promptly observe and reconstruct 
the standard candles for the heavy quark physics, 
the \jpsi and \upsi states, in a quasi-on-line fashion.
Then soon after, with $0.314$ and $3$ pb$^{-1}$ of integrated 
luminosity, respectively, full analyses were performed and published~\cite{cms:jpsi, cms:upsi}.
In both \jpsi and \upsi analyses, the data were collected with a trigger requiring 
the detection of two muons at the hardware level, without any further
selection at higher level\footnote{The coincidence of two muon 
signals, without any explicit \pt requirement, is sufficient 
to maintain the dimuon trigger without prescaling.}, 
the resonant states were reconstructed through their decay
in two opposite sign muons, and the production cross sections 
were measured diffentially in \pt and $y$ intervals.

Concerning the datasets analysed,
quality requirements 
are applied on the status of all the sub--components involved, 
on the reconstructed Primary Vertex (PV), as well as on the reconstructed muons.
The kinematic requirements on the muons are chosen to ensure 
that the trigger and muon reconstruction efficiencies 
are high and not rapidly 
changing within the acceptance window considered:
for the muons in the \jpsi analysis, the minimum \pt ranges from $3.3$ 
GeV in the central $\eta$ region, down to $2.4$ GeV in the endcap regions, 
whereas in the \upsi study, 
one asks for muons with \pt$>3.5$ GeV in the barrel 
and \pt$>2.5$ GeV in the endcaps. 

The detector systems have been aligned and calibrated using LHC 
collision data and cosmic ray 
muons~\cite{bib-magneticfield,bib-material,bib-trackeralignment}. 
Further residual effects are determined by studying the
dependence of the reconstructed \jpsi dimuon invariant-mass 
distribution on the muon kinematics. The transverse momentum is then
corrected for the residual scale, for both analyses, 
the parameters having been evaluated through 
a likelihood fit performed to the peak of the invariant mass shape.
To reduce the combinatorial background, one also requires
the invariant mass of the two muons to lie 
in a certain window centered around the nominal value of the mass for the 
considered resonance.

\subsection{Acceptance and efficiency}
The muon acceptance, defined here as the fraction of detectable 
muons from the meson decay as a function of the dimuon 
transverse momentum and rapidity, 
reflects the finite geometrical coverage of the CMS detector 
and the limited kinematical reach of muon trigger and reconstruction 
systems. 

To compute the acceptance, \jpsi and \upsi simulated events 
have been generated with no cut on 
\pt and  within a pseudorapidity region extending beyond the 
muon detector coverage.
In the signal Monte Carlo (MC) sample,
the meson decays into two muons are generated with the
\textsc{evtgen}~\cite{evtgen} package, including 
effects of final-state radiation, fully simulated and then 
reconstructed with the CMS detector simulation software, 
to assess multiple scattering and finite resolution effects.
In general, the acceptance varies with the resonance's mass, and moreover
the meson polarization strongly influences the
muon angular distributions, expected to change as a function of $p_\mathrm{T}$.  
In order to account for all of this, the acceptance is calculated for
five extreme polarization scenarios~\cite{bib-faccioli}: 
unpolarized, polarized longitudinally and transversely 
with respect to two different reference frames: 
the helicity frame, defined by the
flight direction of the meson in the center-of-mass system of the
colliding beams, and the Collins-Soper
frame~\cite{bib-CollinsSoper}, given by the direction of the
incoming protons direction in the meson's rest frame.
Fig. 1 shows the expected values of the acceptance for 
muons coming from the \jpsi and \upsi decays, as a function 
of the meson's \pt and $y$, for the unpolarized scenario.

\begin{figure}[htb!]
\centering
\includegraphics[width=0.49\textwidth,height=7cm]{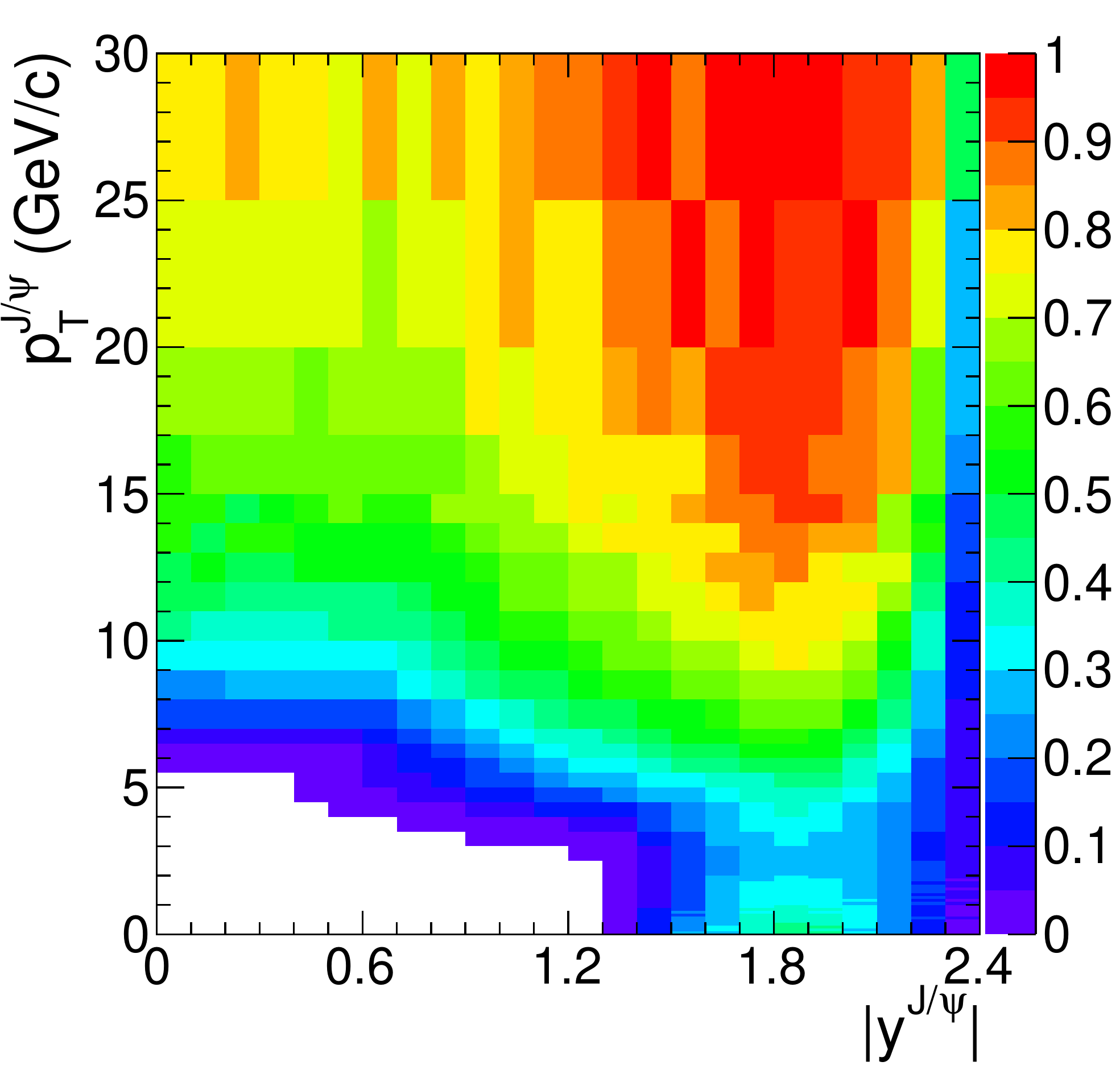}%
\includegraphics[angle=90,width=0.49\textwidth,height=7cm]{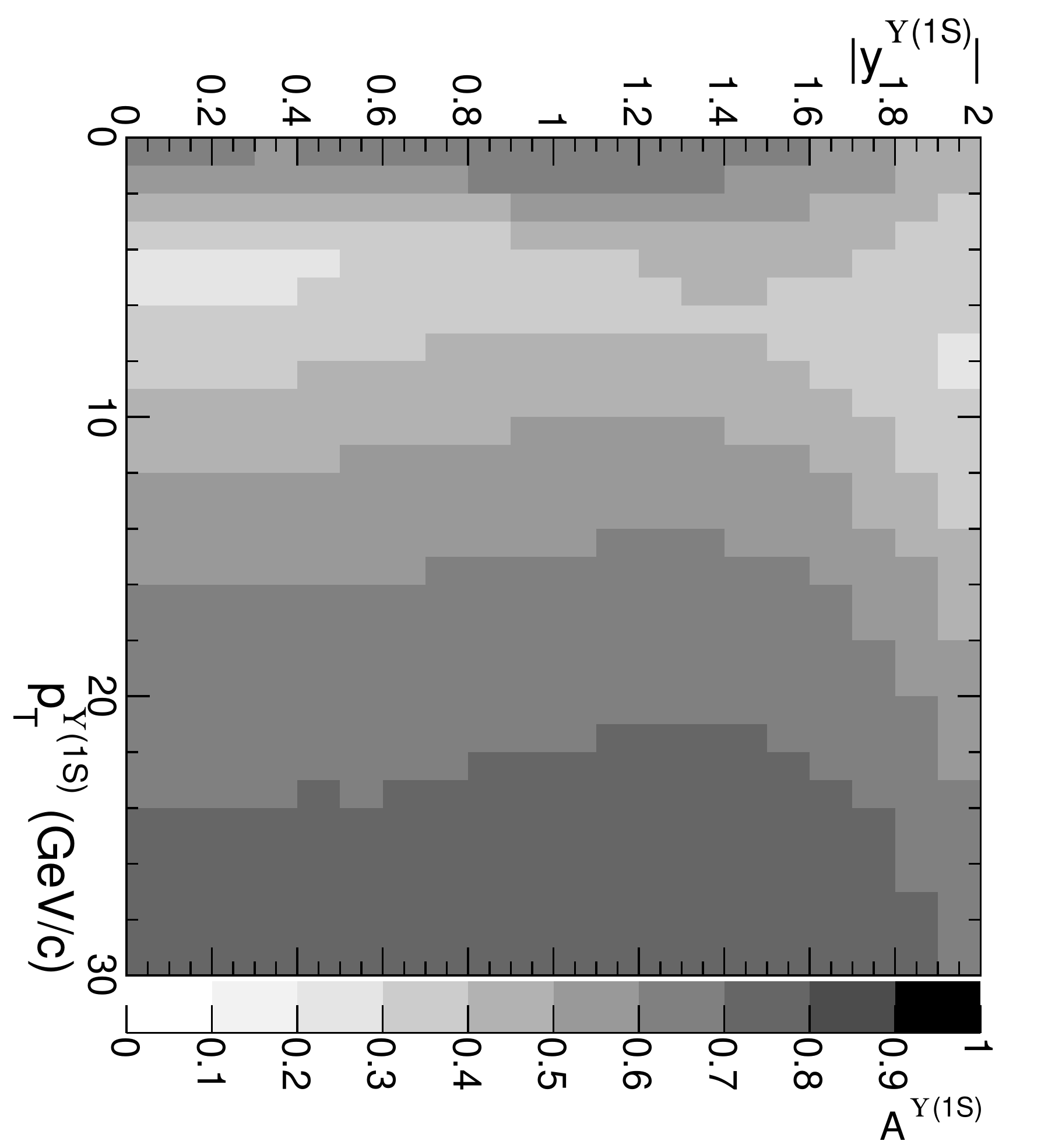}
\small{
\caption{Expected values of acceptance for 
muons coming from \jpsi(left) and \upsi(right) decays, as a function 
of the meson's \pt and $y$, for the unpolarized scenario, as estimated from
simulated events. Note that for \jpsi, the acceptance is mapped in bins of
\pt vs $y$, and in  the opposite way for \upsi.}}
\label{acceptances}
\end{figure}

The total muon efficiency can be factorized into three terms, 
accounting for the trigger efficiency, the 
muon identification efficiency and of good quality track reconstruction, respectively.
All the three components are evaluated with the ``Tag and Probe''
data--driven method~\cite{TnP}, considering a data sample 
selected with looser trigger requirements.
In events with two muon candidates, one
candidate, called the ``tag'', is required to satisfy tighter identification
criteria. The other candidate, called the ``probe'',
is selected with criteria depending on the measured efficiency.
The efficiency to detect a given \jpsi or \upsi event is thus dependent
on the value of the muon-pair kinematic variables.

The tracking efficiency is found to be constant in the momentum
range defined by the acceptance cuts, and it varies only
slightly in the $|\eta|$ plane. The muon identification and trigger
efficiencies have a stronger \pt
 and $|\eta|$  dependence, which
is mapped with a fine granularity (from nine to twelve bins in \pt and five
in $|\eta|$).

\begin{figure}[htb!]
\centering
\includegraphics[angle=90,width=0.34\textwidth]{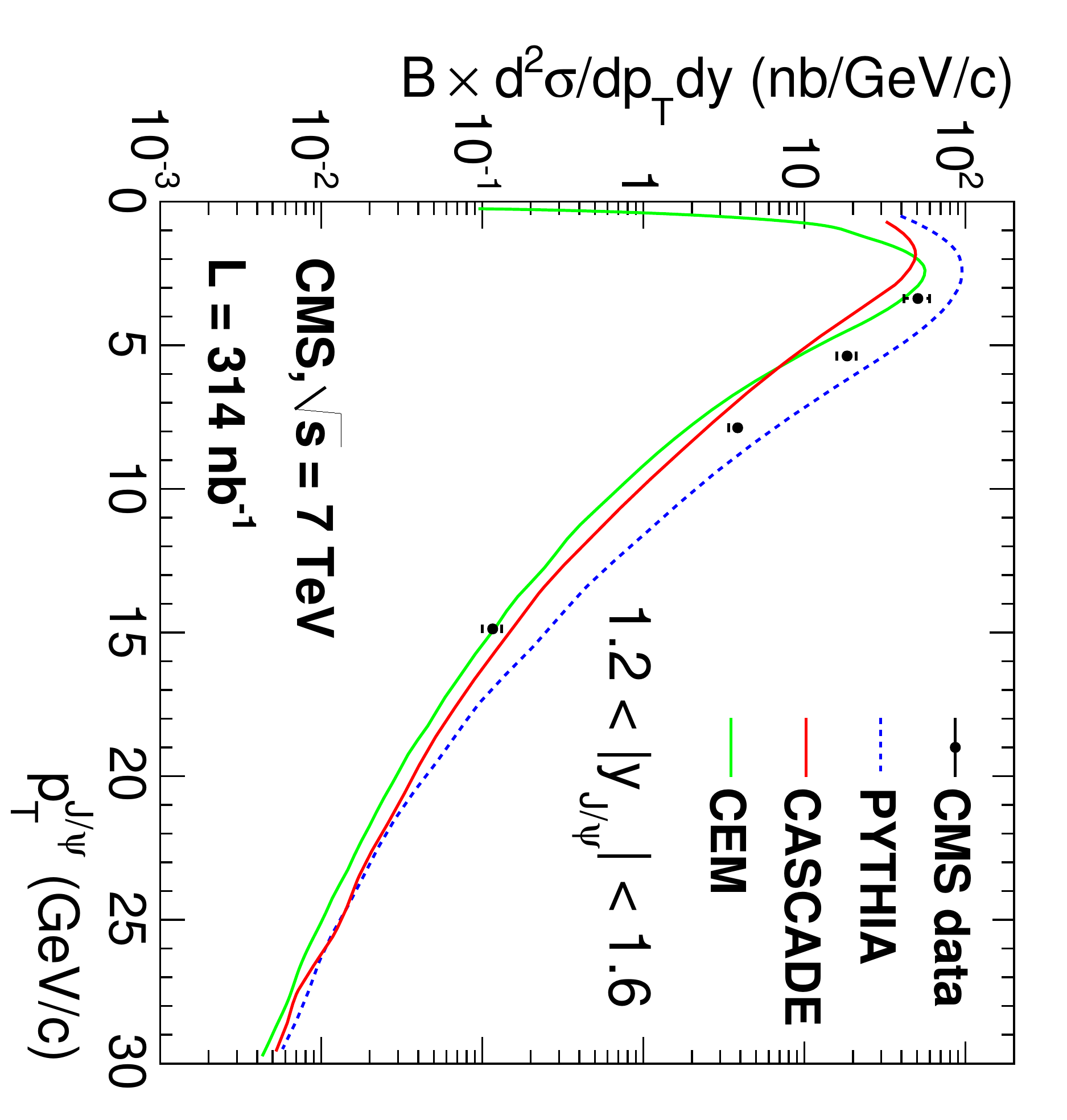}%
\includegraphics[angle=90,width=0.34\textwidth]{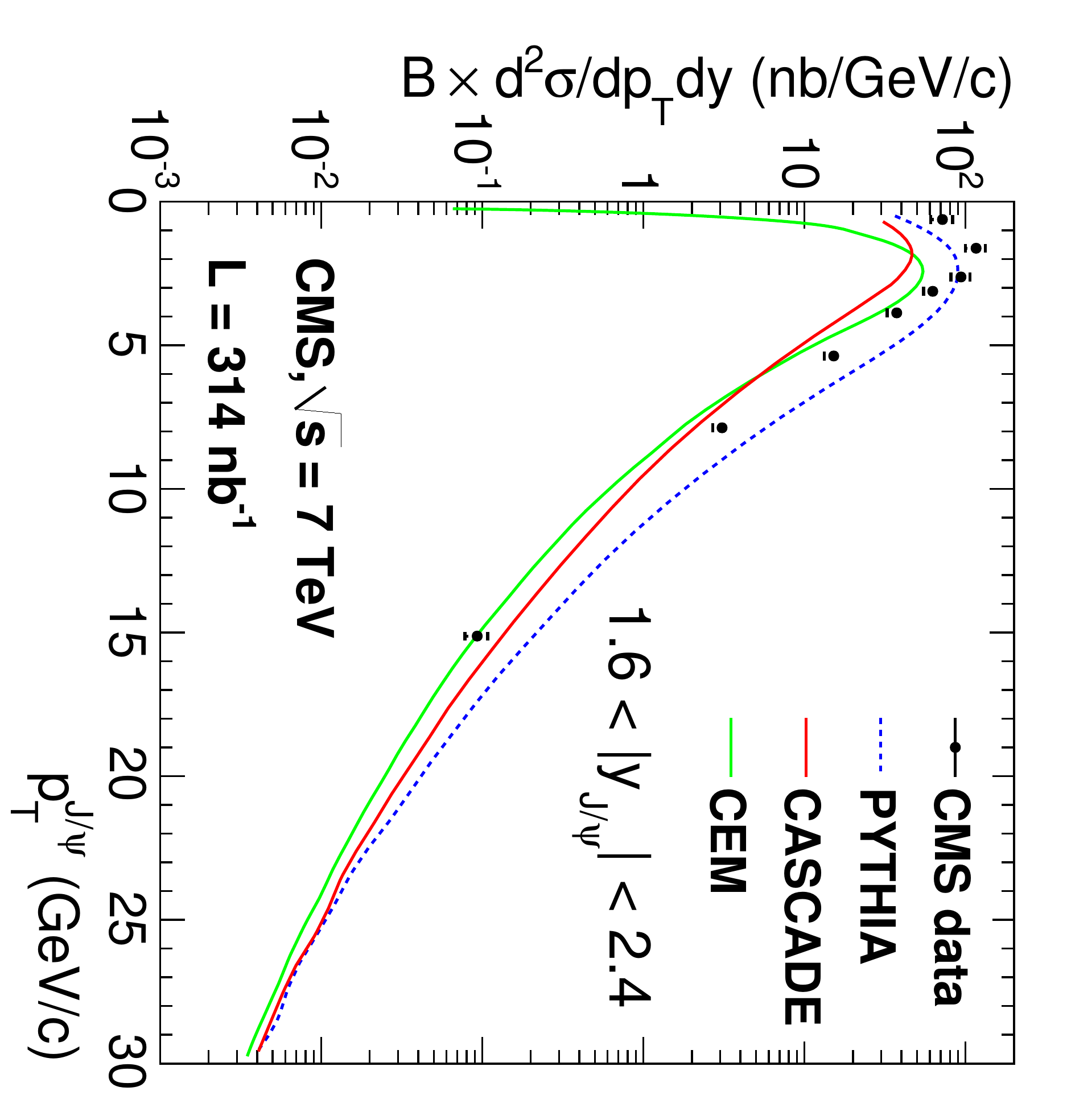}%
\includegraphics[angle=90,width=0.34\textwidth]{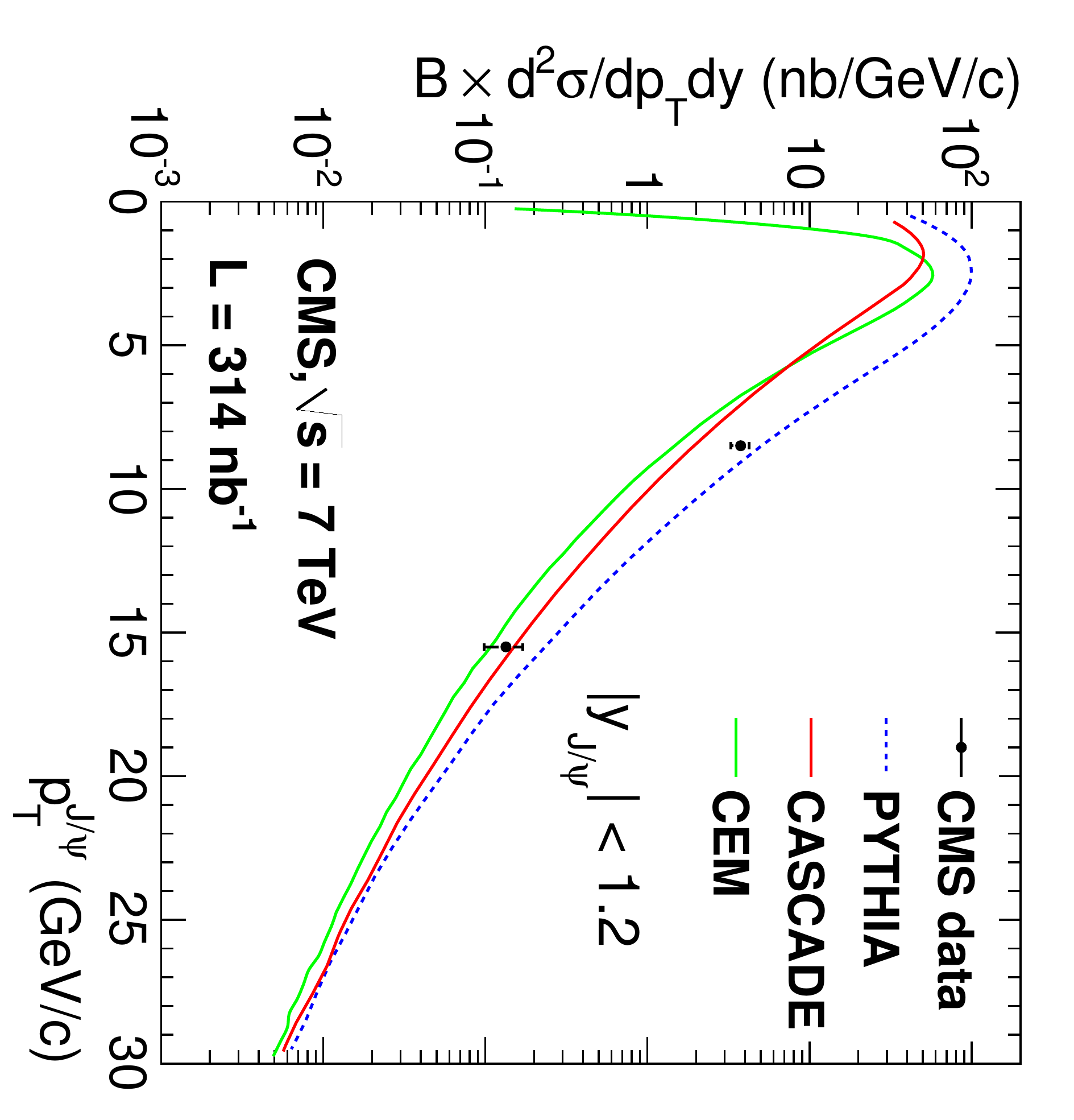}
\caption{\small{Differential prompt \jpsi cross section as function of \pt for the three 
different rapidity intervals reported in the figures, 
for the unpolarized production scenario. The data are
 compared with the available 
teoretical predictions, from Pythia, Cascade MC and Colour Evaporation Model.
The error bars represent the statistical and systematic errors 
added in quadrature. The $11$\% uncertainty due to the luminosity 
determination is not shown and is common to all bins.}
}
\label{fig:jpsixsec}
\end{figure}

\subsection{\jpsi cross-sections measurements}
The differential cross sections are determined from the signals 
yields, obtained directly from a weighted unbinned maximum likelihood fit
to the dimuon invariant-mass spectrum, once corrected for the acceptance
and the overall efficiency. 
In the mass fits, the shape assumed for the signal is
a ``Crystal Ball''~\cite{bib-crystalball} function, which takes into account
the detector resolution as well as the radiative tail from bremsstrahlung. \\
Besides the uncertainty of the luminosity normalization, 
about 11\% for measurements in the first data taking and
analysis period, 
the largest source of systematic uncertainty arises from the 
determination of the muon efficiencies from the data.
It's worth noticing that both the uncertainties reduce with additional data.

The total cross section for 
inclusive \jpsi production, obtained by integrating over \pt between $6.5$ and $30$ GeV
and over rapidity between $-2.4$ and $2.4$, in the unpolarized production hypothesis, gives:
\begin{eqnarray*}
\sigma(pp\rightarrow J/\psi~X)\cdot\mbox{BR}(J/\psi\to\mu^+\mu^-) = (97.5 \pm 1.5\mbox{(stat)} \pm 3.4\mbox{(syst)} \pm 10.7\mbox{(lumi)}) \mbox{ nb}
\end{eqnarray*}

The inclusive \jpsi decays include a prompt and a non--prompt component.
Figure~\ref{fig:jpsixsec} shows the prompt differential cross section
$\frac{d^2\sigma}{d p_{\mathrm{T}} dy} \cdot BR(J\psi \rightarrow \mu^+\mu^-)$
in three rapidity ranges, together with statistical and systematic uncertainties 
(except the luminosity one), added in quadrature. 
The prompt \jpsi differential production cross sections
have been compared with teoretical predictions from Pythia~\cite{bib-PYTHIA}, Cascade~\cite{cascade} Event Generator and the Colour Evaporation Model (CEM)~\cite{ColorEvaporationModel,Halzen:1977rs,Fritzsch:1977ay,Gluck:1977zm,Barger:1979js}. 
These calculations include contributions to the prompt \jpsi yield
due to feed-down decays from heavier charmonium states ($\chi_c$ and
$\psi(2S)$) and can, therefore, be directly compared to the measured
data points, as shown in Fig.~\ref{fig:jpsixsec}\footnote{Since the measurements 
of prompt \jpsi include a significant contribution from feed-down decays, 
of the order of $30$\%~\cite{Ma:2010vd, Faccioli:2008ir}, 
it is not possible to compare the prompt measurement with the predictions 
of models such as the Colour-Singlet Model (including higher-order 
corrections)~\cite{Lansberg:2008gk}
or the LO NRQCD model (which includes singlet and octet 
components),
available only for the \emph{direct} \jpsi.}.
At forward rapidity and low \pt the calculations underestimate the measured yield.

\subsection{Non-prompt \jpsi cross section}\label{sec:bfraction}
The estimation of non-­‐prompt \jpsi
coming from b--hadron decays
can be performed by discriminating \jpsi mesons produced
away from the $pp$ collision vertex, that means observing the
distance between the dimuon vertex and the primary vertex 
in the plane orthogonal to the beam line.
To determine the  non-prompt \jpsi production cross-section, an unbinned maximum-likelihood fit in each \pt and rapidity bin was performed
to the distributions of the invariant mass of the muon pair and to 
the quantity
$l_{J/\psi}=L_{xy}\cdot m_{J/\psi}/{p_{\mathrm{T}}}$ 
where $L_{xy}$ is the most probable transverse decay length in the laboratory frame. 

\begin{figure}[h!]
\centering
\includegraphics[width=0.49\textwidth]{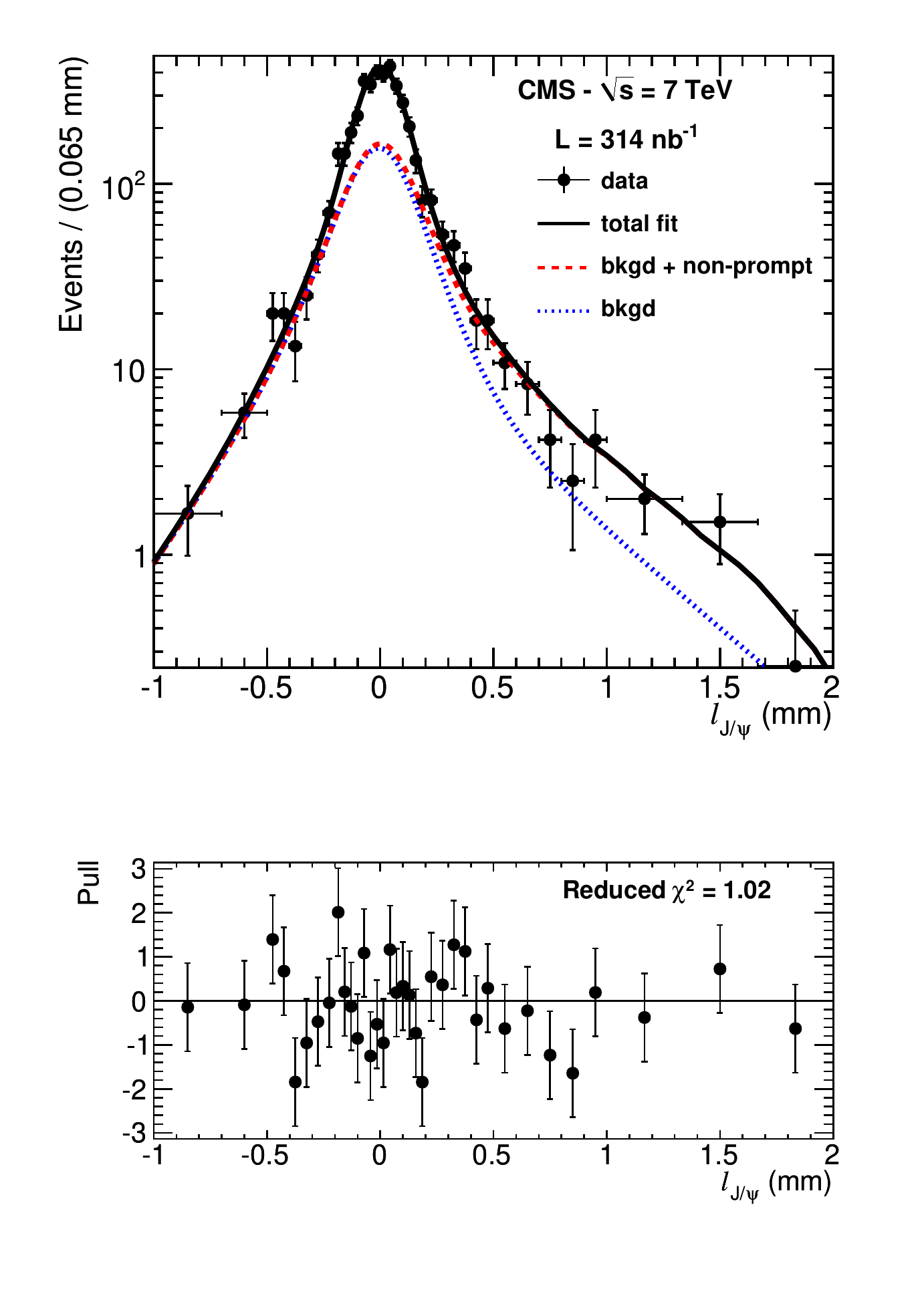}%
\includegraphics[width=0.49\textwidth]{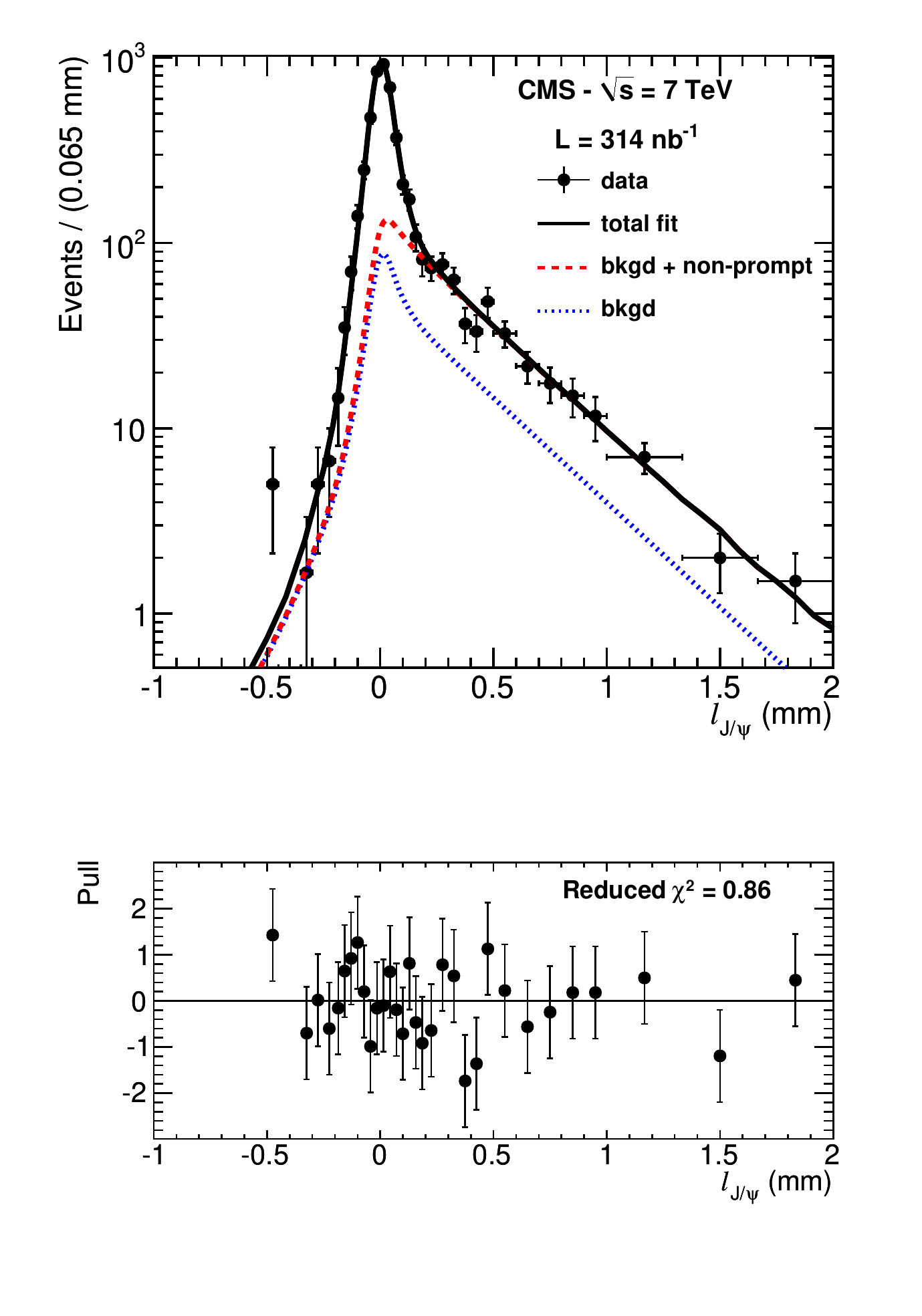}
\caption{
\small{ Projection
of the two--dimensional
likelihood fit
(in mass and $l_{J/\psi}$) to the $l_{J/\psi}$
dimension for bins $2<p_{\mathrm{T}}<4.5$ GeV,  $1.2<|y|<1.6$ (left) and
$6.5<p_{\mathrm{T}}<10$ GeV, $1.6<|y|<2.4$ (right), with their pull distributions (bottom).}}
\label{fig:distr_lxy_data}
\end{figure}

\begin{figure}[h!]
\centering
\includegraphics[angle=90,width=0.34\textwidth]{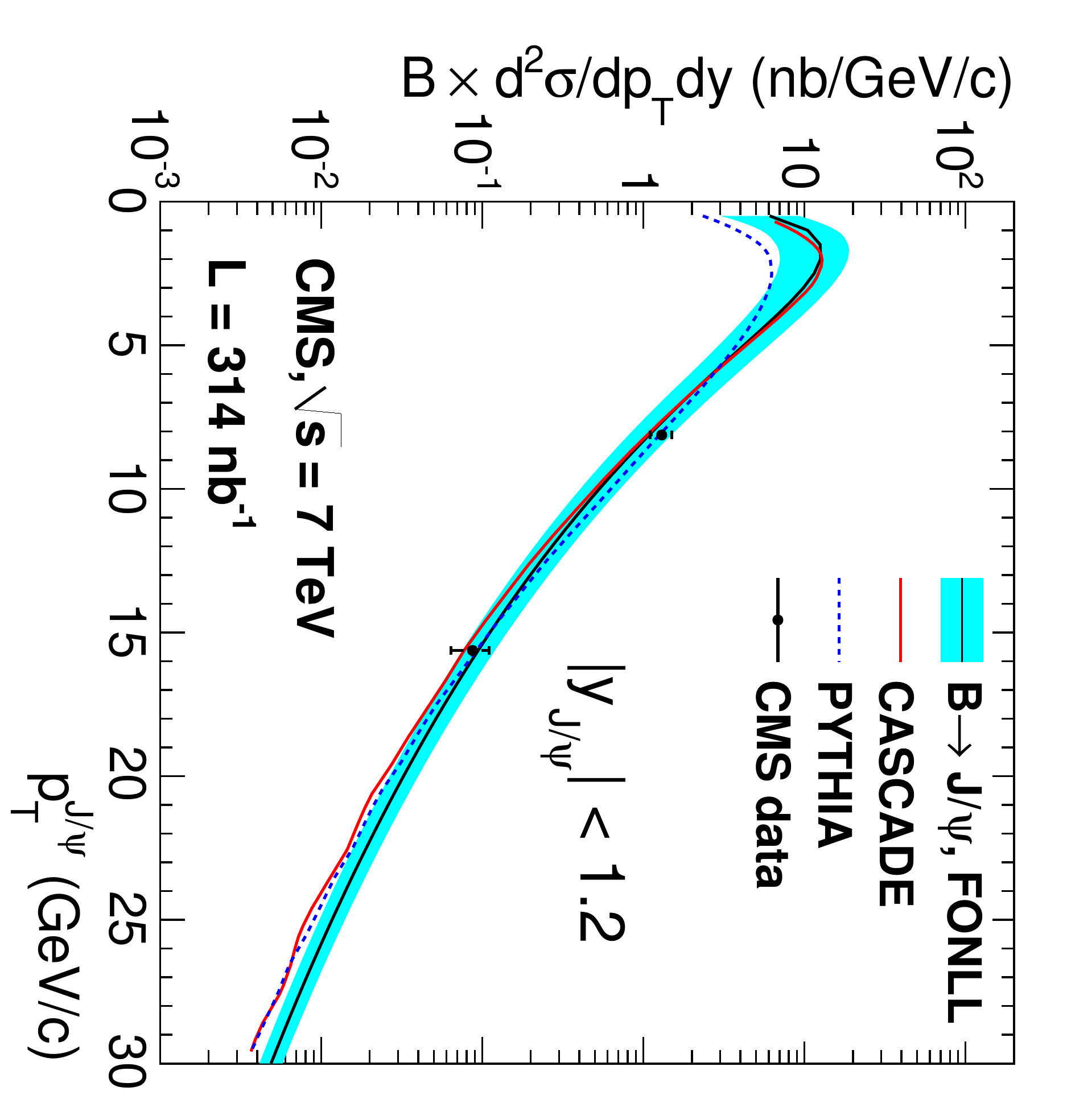}%
\includegraphics[angle=90,width=0.34\textwidth]{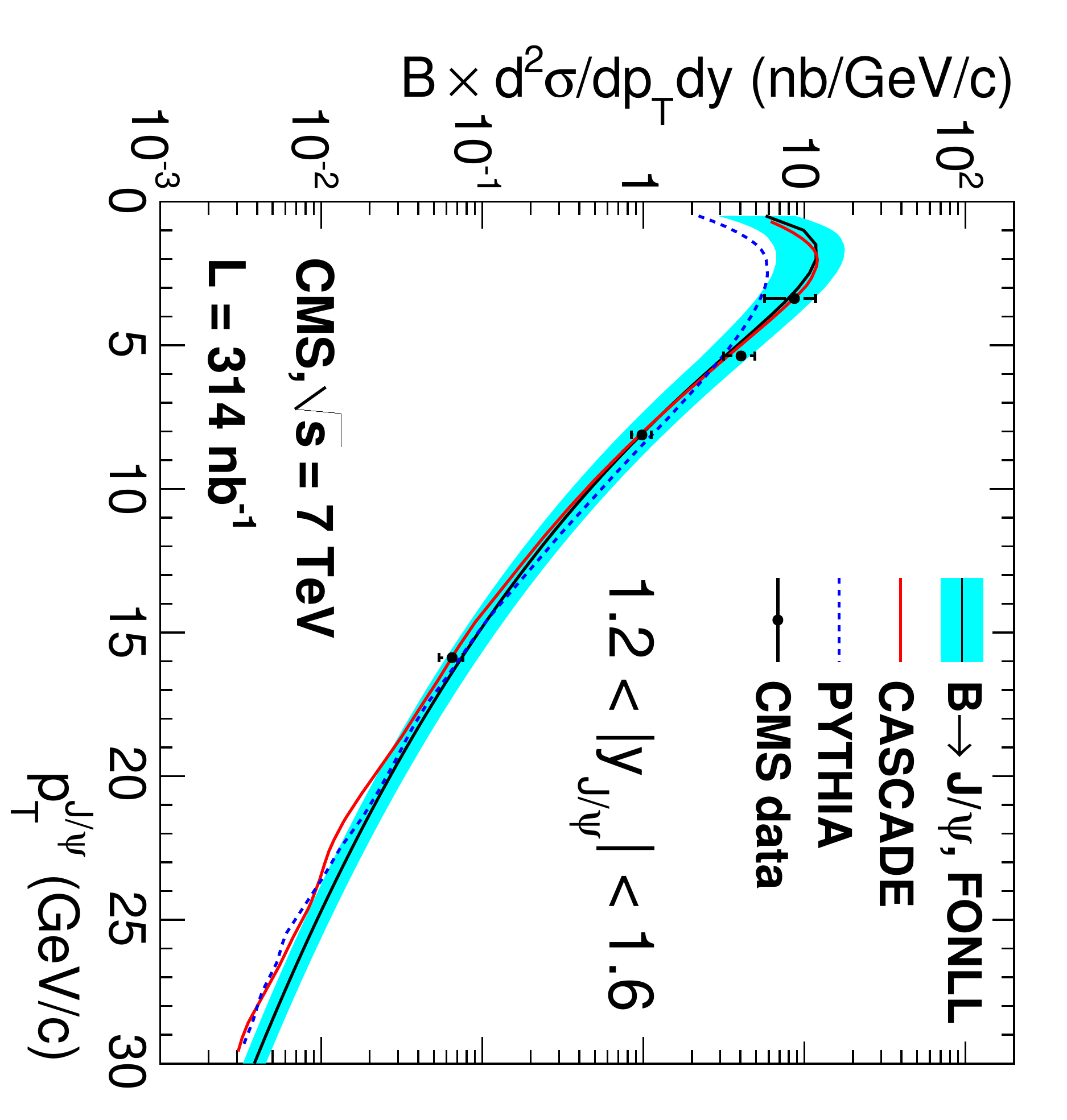}%
\includegraphics[angle=90,width=0.34\textwidth]{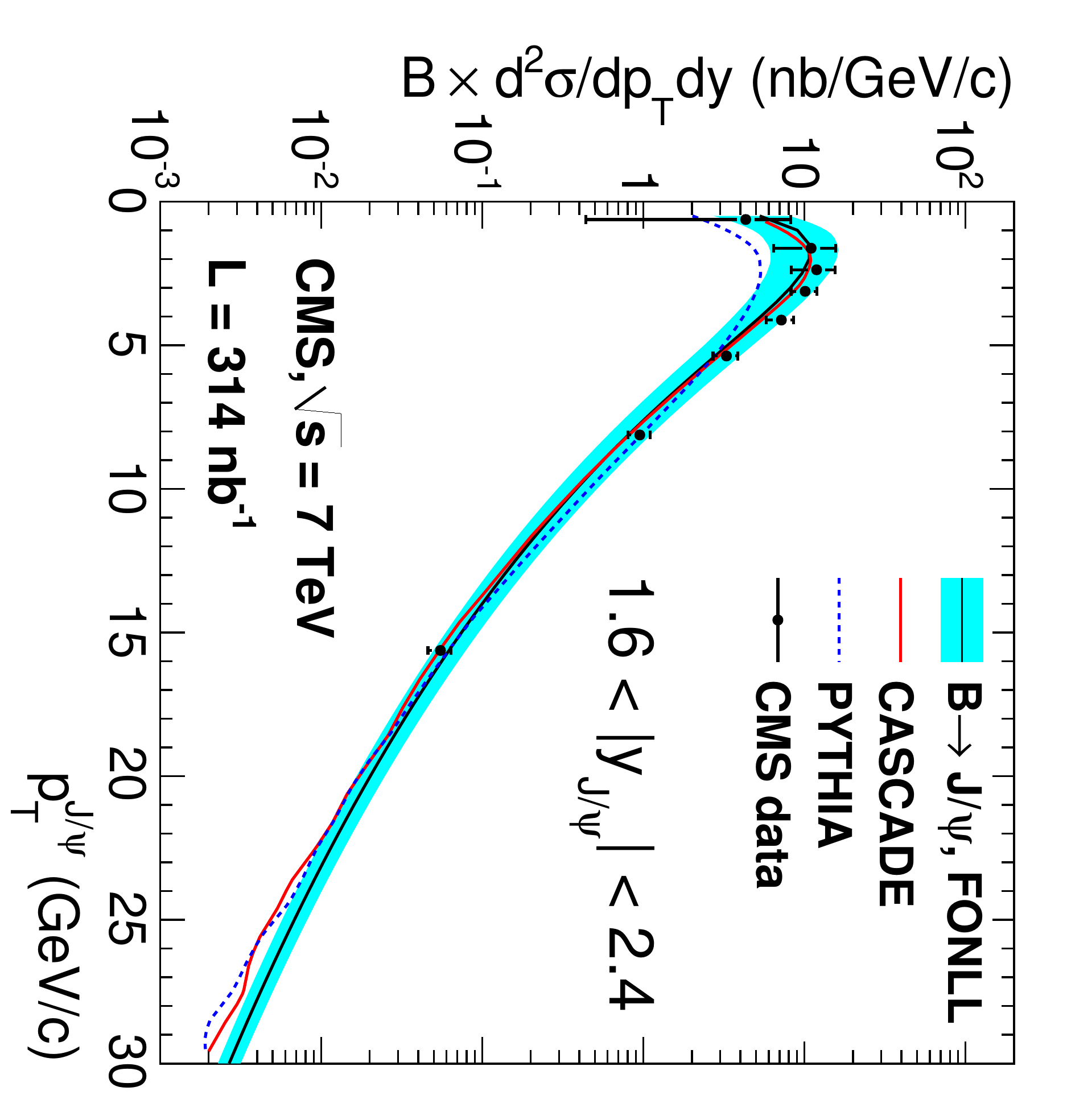}
\caption{
\small{Differential non-prompt \jpsi production cross section, as a
function of \pt for three different rapidity intervals.
The data points are compared with three different models, using the PYTHIA curve to
calculate the abscissa where they are plotted~\cite{wyatt}.}}
\label{fig:theorynonprompt}
\end{figure}

Fig.~\ref{fig:distr_lxy_data} shows the projection of the likelihood fits
in two sample bins.
The non-prompt \jpsi differential production cross sections have also been compared with calculations
made with the Pythia and CASCADE Monte Carlo generators, and in the FONLL
framework~\cite{FONLL}. The results are presented in
Fig.~\ref{fig:theorynonprompt}, showing a good agreement with the calculations.


Different sources of systematic errors contribute to the total uncertainty of the
extraction of the non--prompt component, and in general they 
depend on the considered $y$ interval. In rough order of relevance, they are:
\begin{itemize}
\item primary vertex estimation;
\vspace{-0.2cm}
\item decay length resolution function;
\vspace{-0.2cm}
\item background fit;
\vspace{-0.2cm}
\item residual misalignments in the tracker;
\vspace{-0.2cm}
\item b-hadron lifetime model;
\vspace{-0.2cm}
\item efficiencies for prompt and non--prompt \jpsi.
\end{itemize}

\subsection{\upsi cross section measurements}

The \upsi states $1$S, $2$S and $3$S were observed with 
evident signals above the background in the $3$ pb$^{-1}$ of data analyzed. 
The mass resolution provided by the full CMS detector 
reconstruction is $\sim 70$ MeV for $ \eta(\mu)< 1.0$.
The yields are extracted simultaneously with a maximum likelihood fit in \pt and $y$ intervals, and the
double differential cross sections in \pt and $y$  are measured after having corrected the data for 
the acceptance and efficiency as for the \jpsi analysis.
The measured values of the $\Upsilon(nS)$ integrated production cross sections 
for the rapidity range $|y|<2$ are:

\small{
\begin{eqnarray*}
\centering
    \sigma(p~p \rightarrow \Upsilon(1S) X ) \cdot B(\Upsilon (1S) \rightarrow \mu^+
    \mu^-) & = & (7.37 \pm 0.13 (\mathrm{stat})^{+0.61}_{-0.42} (\mathrm{syst})\pm 0.81 (\mathrm{lumi}))\mathrm{nb};  \\
    \sigma(p~p \rightarrow \Upsilon(2S) X ) \cdot B(\Upsilon(2S) \rightarrow \mu^+ 
    \mu^-) & = & (1.90 \pm 0.09 (\mathrm{stat})^{+0.20}_{-0.14} (\mathrm{syst})\pm 0.24 (\mathrm{lumi}))\mathrm{nb};  \\
    \sigma(p~p \rightarrow \Upsilon(3S) X ) \cdot B(\Upsilon(3S) \rightarrow \mu^+
    \mu^-) & = & (1.02 \pm 0.07 (\mathrm{stat})^{+0.11}_{-0.08} (\mathrm{syst}) \pm 0.11 (\mathrm{lumi}))\mathrm{nb}. 
\end{eqnarray*}
}

The \upsi$(1S)$ and \upsi$(2S)$ measurements include 
feed-down from higher-mass states, such as the $\chi_b$ family and the \upsi$(3S)$.
These measurements assume unpolarized  production.
Assumptions of fully-transverse or fully-longitudinal polarizations 
change the cross sections by about 20\%.

\begin{figure}[h!]
\centering
\includegraphics[angle=90,width=0.49\textwidth]{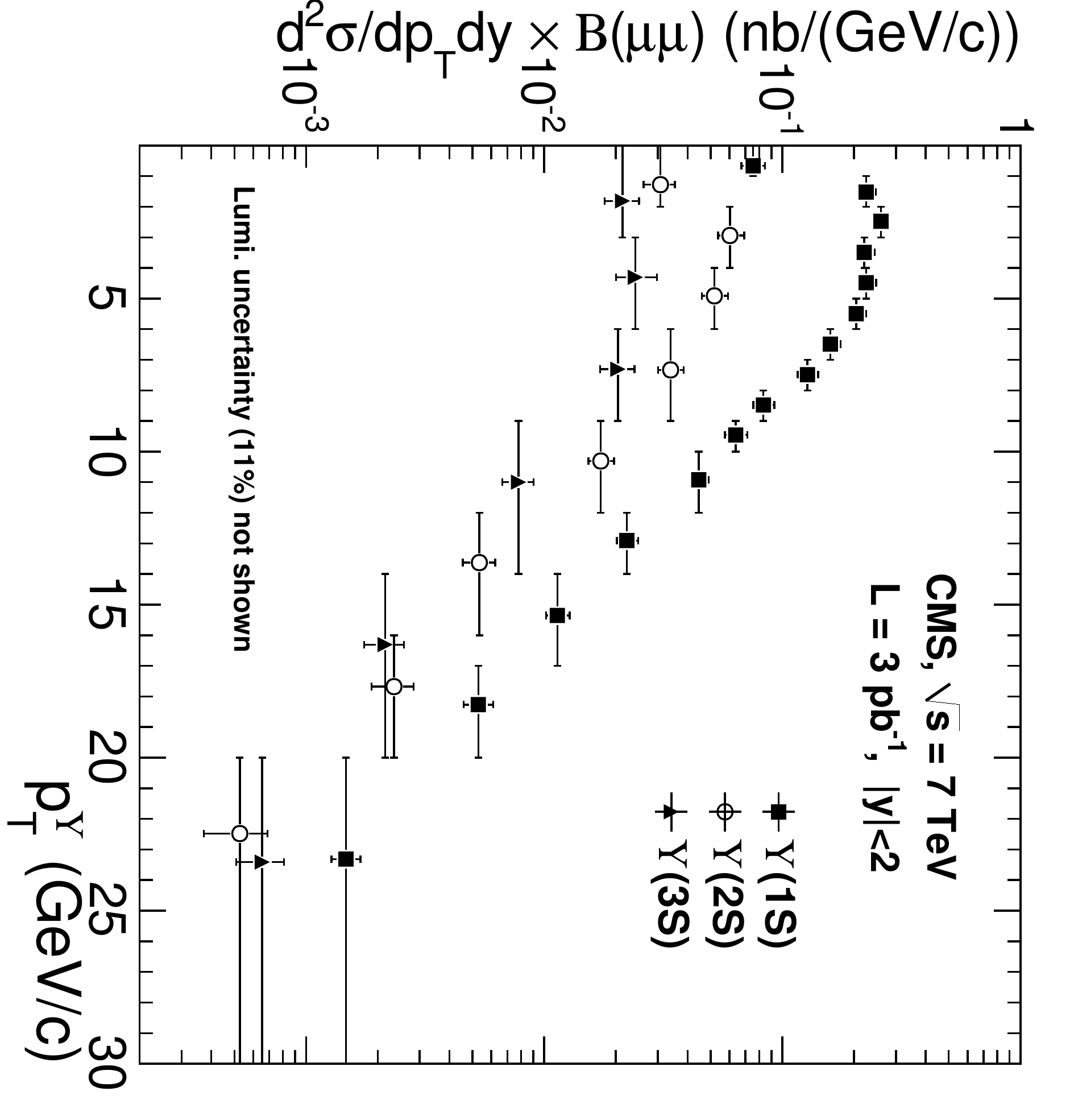}%
\includegraphics[angle=90,width=0.49\textwidth]{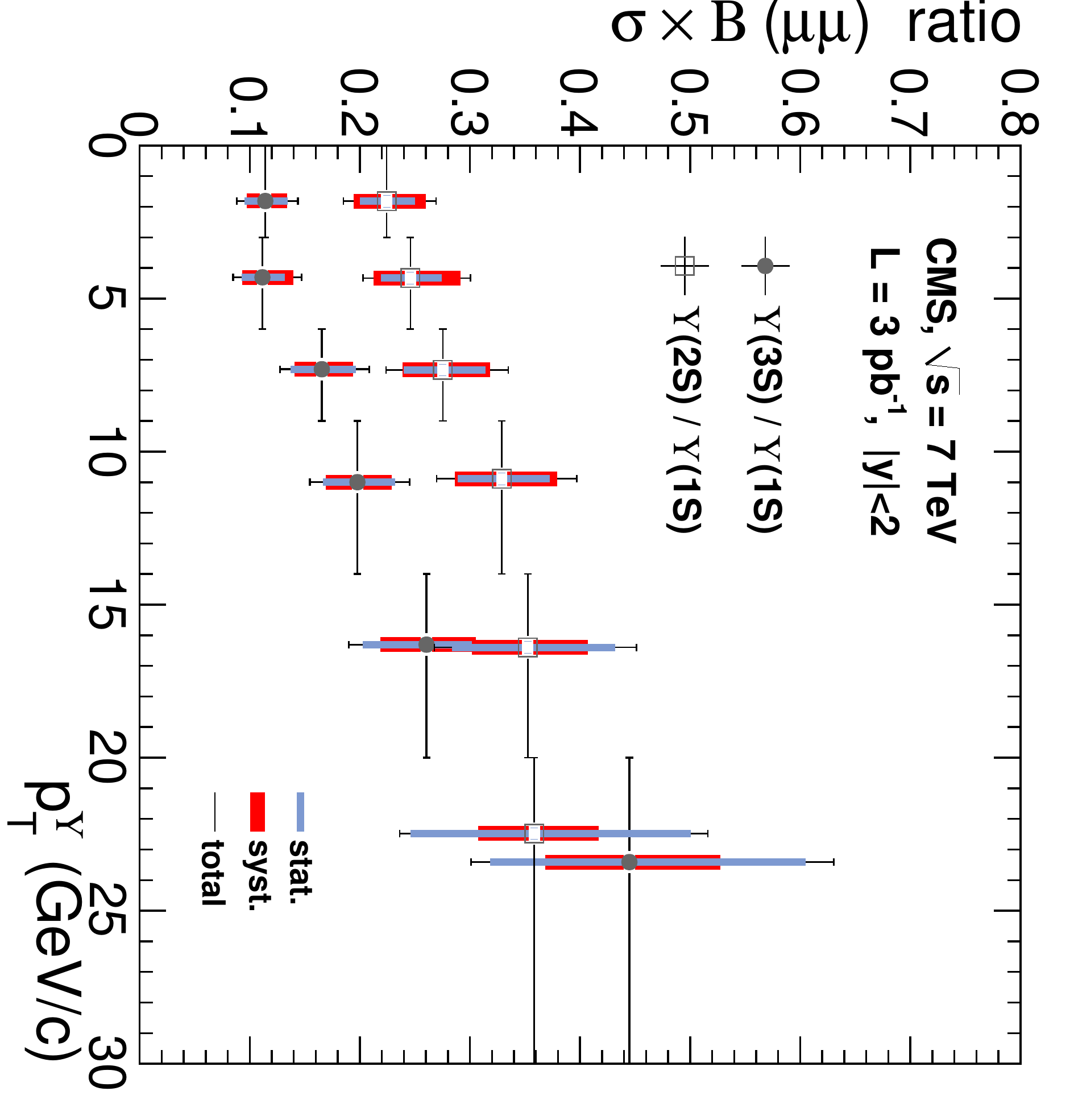}
\caption{
\small{Differential production cross section of $\Upsilon(1\mathrm{S})$, 
$\Upsilon(2\mathrm{S})$ and $\Upsilon(3\mathrm{S})$  states, as a
function of mesons' $p_\mathrm{T}$, on the left hand side, and the cross section 
ratio of $\Upsilon(2\mathrm{S})$  and $\Upsilon(3\mathrm{S})$  
states relative to the ground state, on the right hand side.}}
\label{fig:upsilame}
\end{figure}

The results are shown in Fig. \ref{fig:upsilame}, that presents
the differential production cross sections for the $\Upsilon(nS)$ states
as a function of the meson's $p_\mathrm{T}$, and the relative ratio 
between the 2S and 3S states over the ground $1$S state. 
The fraction of $2$S and $3$S with respect to the $1$S clearly increases with $p_\mathrm{T}$,
The measured values are found to be in good agreement with previous Tevatron findings~\cite{d0:upsipol, cdf:upsipol}.
The normalized \pt spectrum prediction from PYTHIA is consistent
with the measurements, while it overestimates the integrated cross section 
by about a factor of two. 

\section{Result from CDF Experiment on $\Upsilon(1\mathrm{S})$ polarization }
The CDF Collaboration has a long record of studies about quarkonia physics.
Knowledge of \jpsi and \upsi production 
mechanisms has increased in the past decades with new data and with progress in theory.
Still, almost $40$ years after the \jpsi discovery, a comprehensive view of quarkonium 
production mechanisms and polarization is still missing, as already mentioned in Sec.~\ref{ref:intro}.

\begin{figure}[h!]
\centering
\includegraphics[width=1.\textwidth]{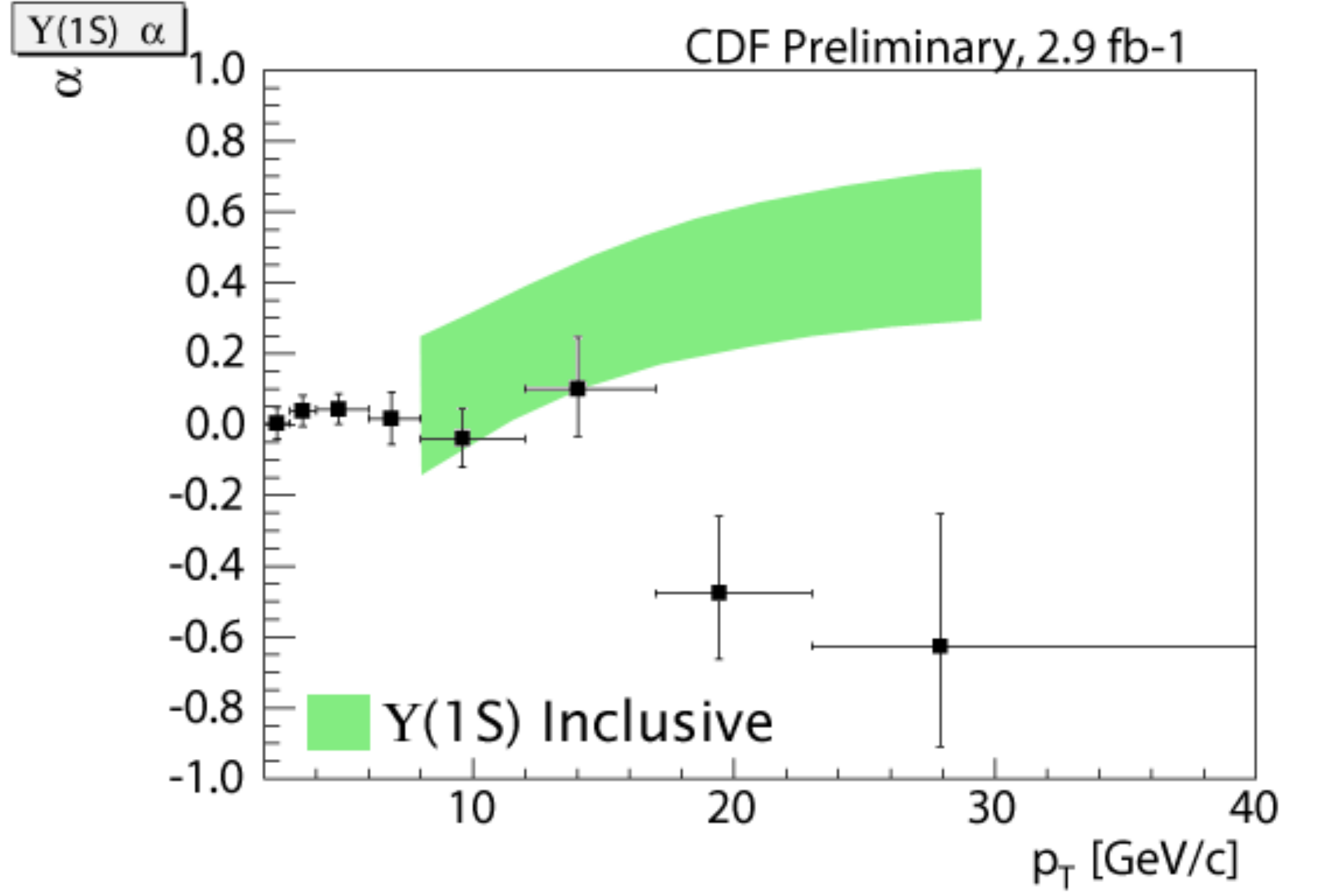}
\caption{
\small{\upsi polarization as a function of the parameter $\alpha$, the polar angle 
of the positive muon in the s-channel helicity frame, as measured by the CDF
Collaboration over 2.9 fb$^{-1}$ of \pp collision data, 
compared with the predictions of the NRQCD calculations, represented
by the colored band.}}
\label{fig:CDFupsi}
\end{figure}

The most recent results about the \upsi polarization study
at CDF~\cite{cdf:latestUpsi},
presented for the first time at this conference, are summarized here. 
The measurements were performed using $2.9$ fb$^{-1}$ of \pp collisions data at 
$1.96$ TeV, making use of a template method, in which fully transverse and fully longitudinal Monte Carlo samples 
are generated and subjected to detector acceptance and efficiency effects, and then 
iteratively re-weighted to match the data \pt distributions. 
Based on the shapes of the polar distributions of the positive 
muon in the $s$-channel helicity frame, 
the polarization parameter is determined by matching 
a polarization-weighted combination of templates to data in different \mbox{
$p_{\mathrm{T}}(\Upsilon)$} bins.
Events are selected using a mass fit, and backgrounds are evaluated using mass sidebands.
The result is presented in Fig.~\ref{fig:CDFupsi}, that represents the \upsi polarization
in terms of the parameter $\alpha$, the polar angle 
of the positive muon in the s-channel helicity frame. Its values range
from $-1$, for a fully longitudinal polarized $\Upsilon$, to $+1$, for a fully transverse polarization.
The \upsi are found to be unpolarized at low $p_\mathrm{T}$, 
before developing a strongly longitudinal polarization at high $p_\mathrm{T}$, 
in obvious contrast with the NRQCD prediction. 
It is worth to mention that these findings are consistent with those from CDF Run I~\cite{cdf:RunI} and in disagreement with D0 run II ones~\cite{d0:RunII}.

\section{Other beauty hadrons}

\subsection{Exlusive prodution of B-mesons}
The high production cross section of \bb pairs at the LHC energy, together with
the good performance of the CMS detector, made possible the observation of 
B--hadrons such as B$^+$, B$^0$ and B$_{\mathrm{s}}$ rather 
early in the first year of data--taking,
where the data samples used for the full analysis have been $5.8$ pb$^{-1}$ for the former~\cite{cms:B+}, and
$40$ pb$^{-1}$ for the latter two~\cite{cms:B0, cms:Bs}.
The decay channels through which 
the \mbox{B--mesons} have been reconstructed are:
\begin{eqnarray}\label{processes}
B^+ & \rightarrow & J/\psi ~ K^+ \\
B^0 & \rightarrow & J/\psi ~ K_s (\rightarrow \pi^+\pi^-)  \nonumber \\
B_{\mathrm{s}} & \rightarrow & J/\psi ~ \Phi (\rightarrow K^+K^-)  \nonumber
\end{eqnarray}

\begin{figure}[h!]
\centering
\includegraphics[width=0.9\textwidth]{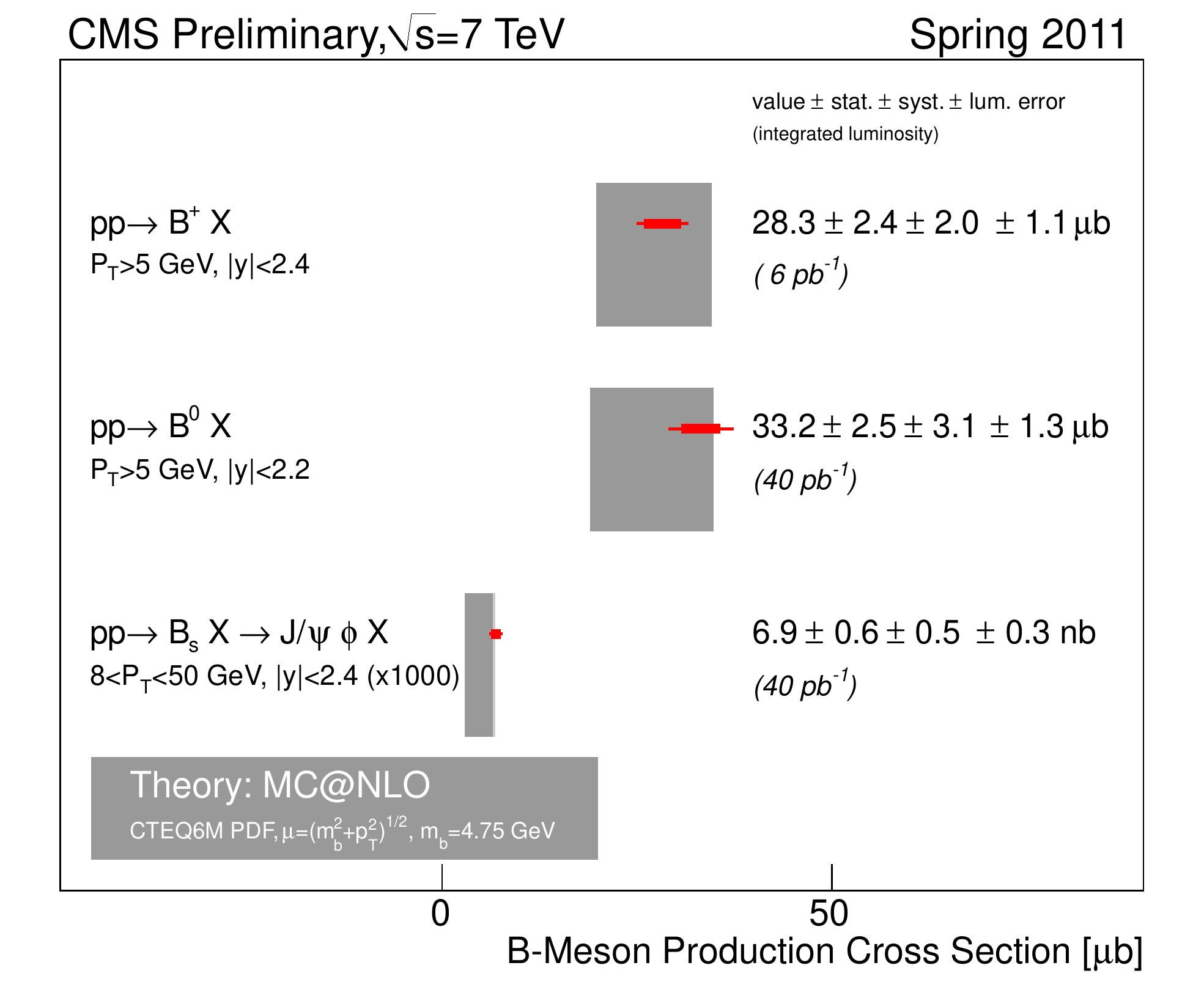}
\caption{
\small{Summary of B meson cross section measurements performed by CMS with 7 TeV  $pp$ collisions at LHC. 
The inner error bars of the data points correspond to the statistical uncertainty, 
while the outer (thinner) error bars correspond to the 
quadratic sum of statistical and systematic uncertainties. 
The outermost brackets correspond to the total error, including a luminosity uncertainty 
which is also added in quadrature. Theory predictions are coming from the MC@NLO.}
}
\label{fig:bHadr}
\end{figure}

The \jpsi is always reconstructed via its decay to two opposite 
sign muons, which are also supposed to fire the di--muon trigger 
used to collect the data sample.
Fig.~\ref{fig:bHadr} summarizes the results for the  
cross section measurements for the channels in (\ref{processes}),
showing also the prediction of the MC@NLO.

\begin{figure}[h!]
\centering
\includegraphics[width=0.49\textwidth]{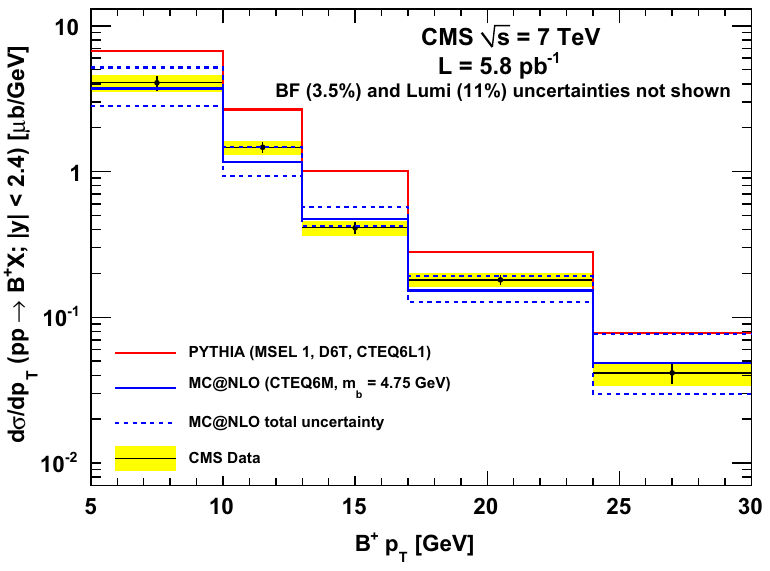}%
\includegraphics[width=0.49\textwidth]{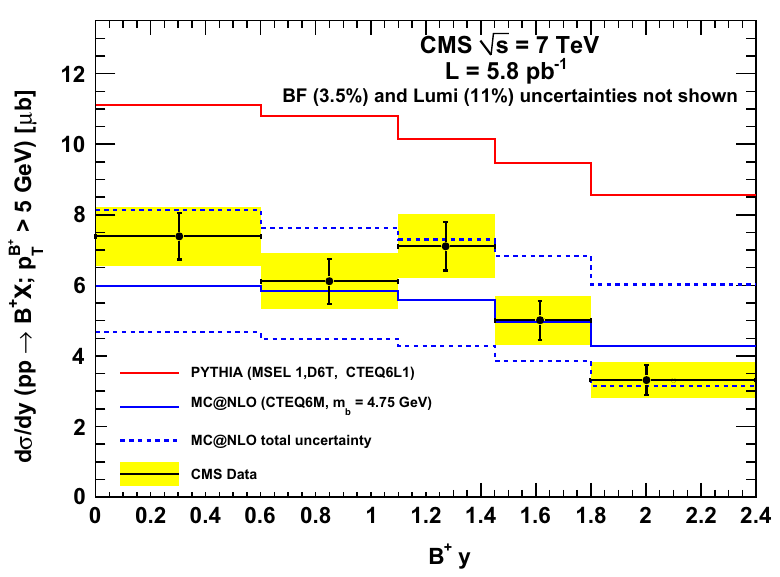}\\
\includegraphics[width=0.49\textwidth]{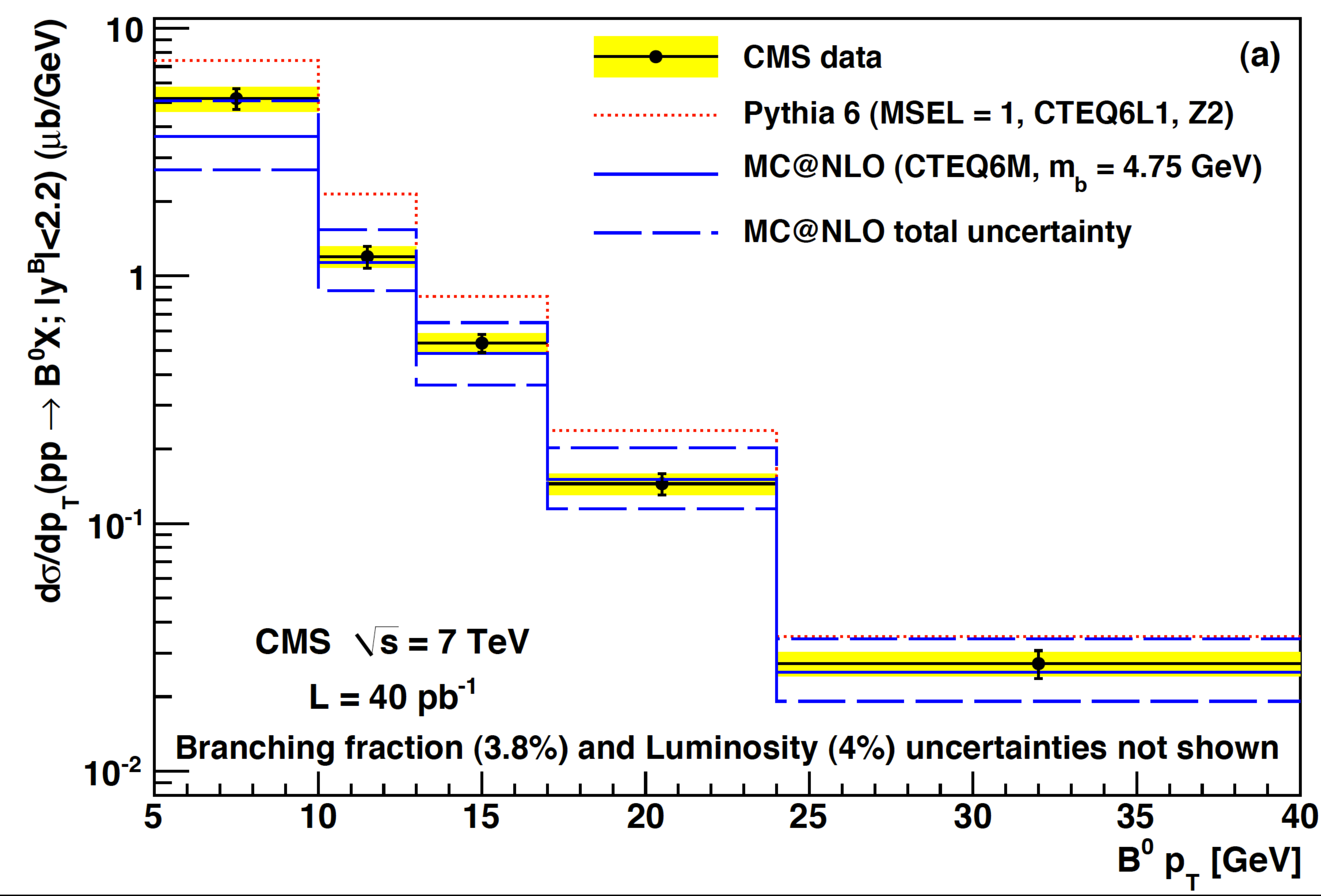}%
\includegraphics[width=0.49\textwidth]{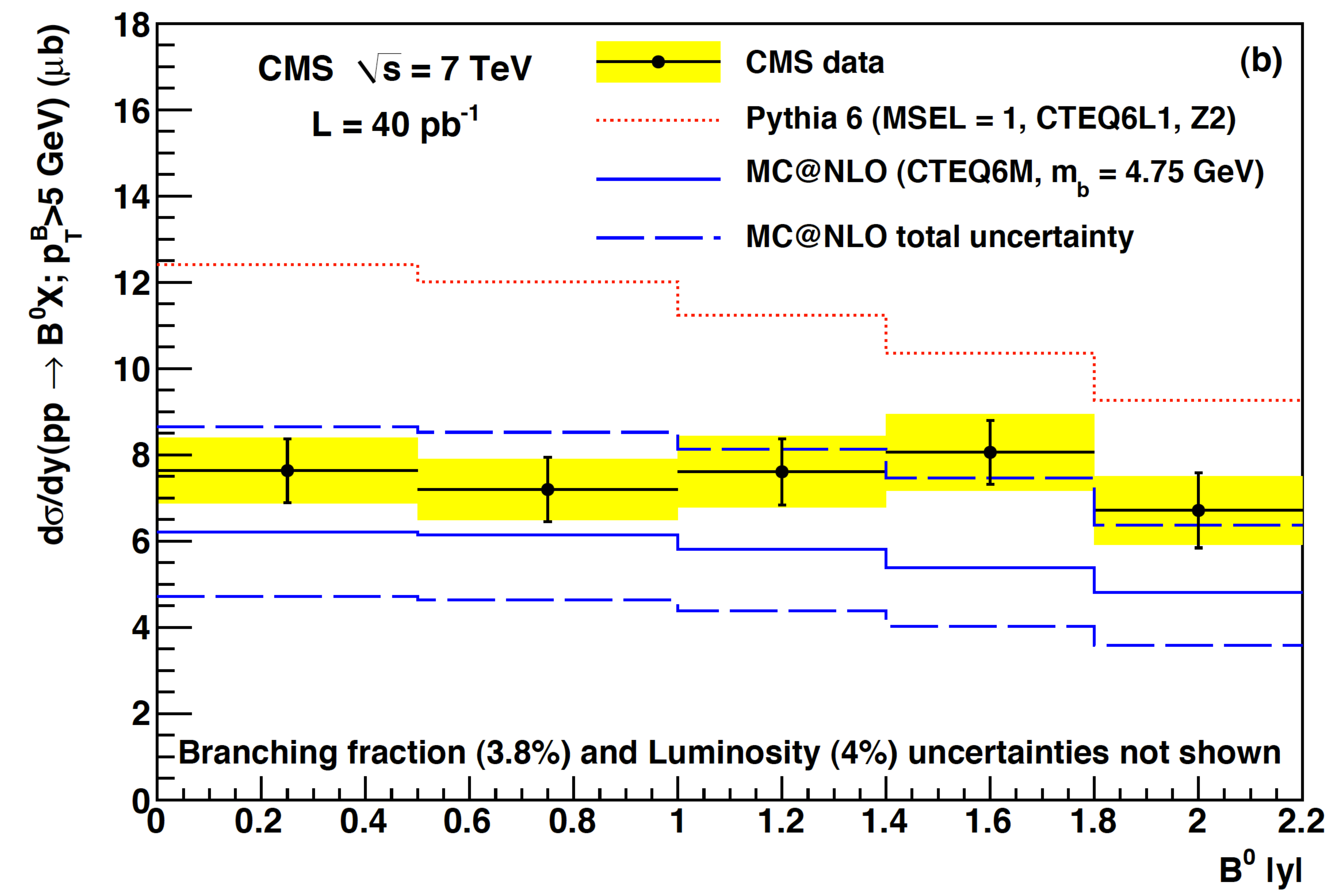}\\
\includegraphics[width=0.51\textwidth]{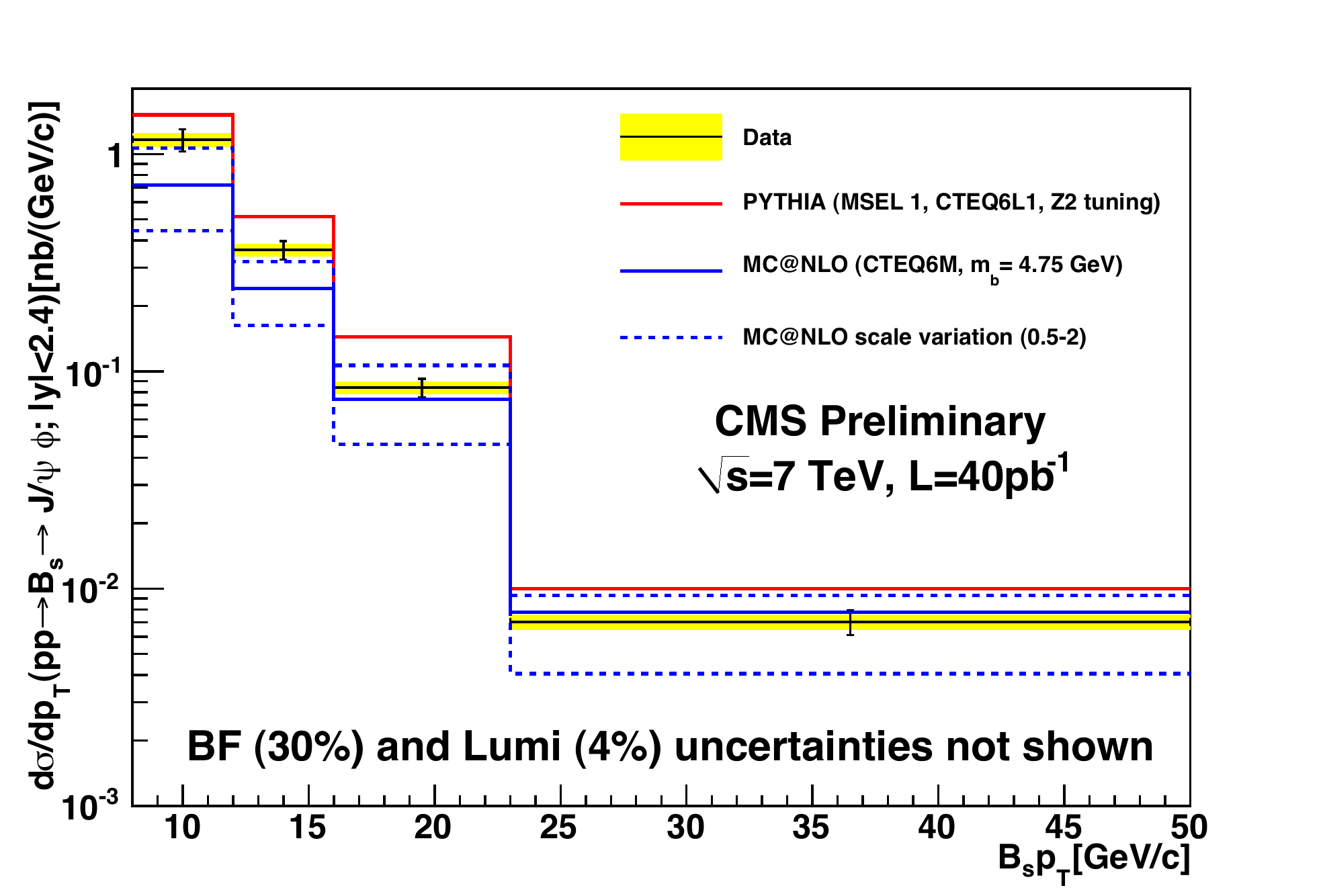}%
\includegraphics[width=0.51\textwidth]{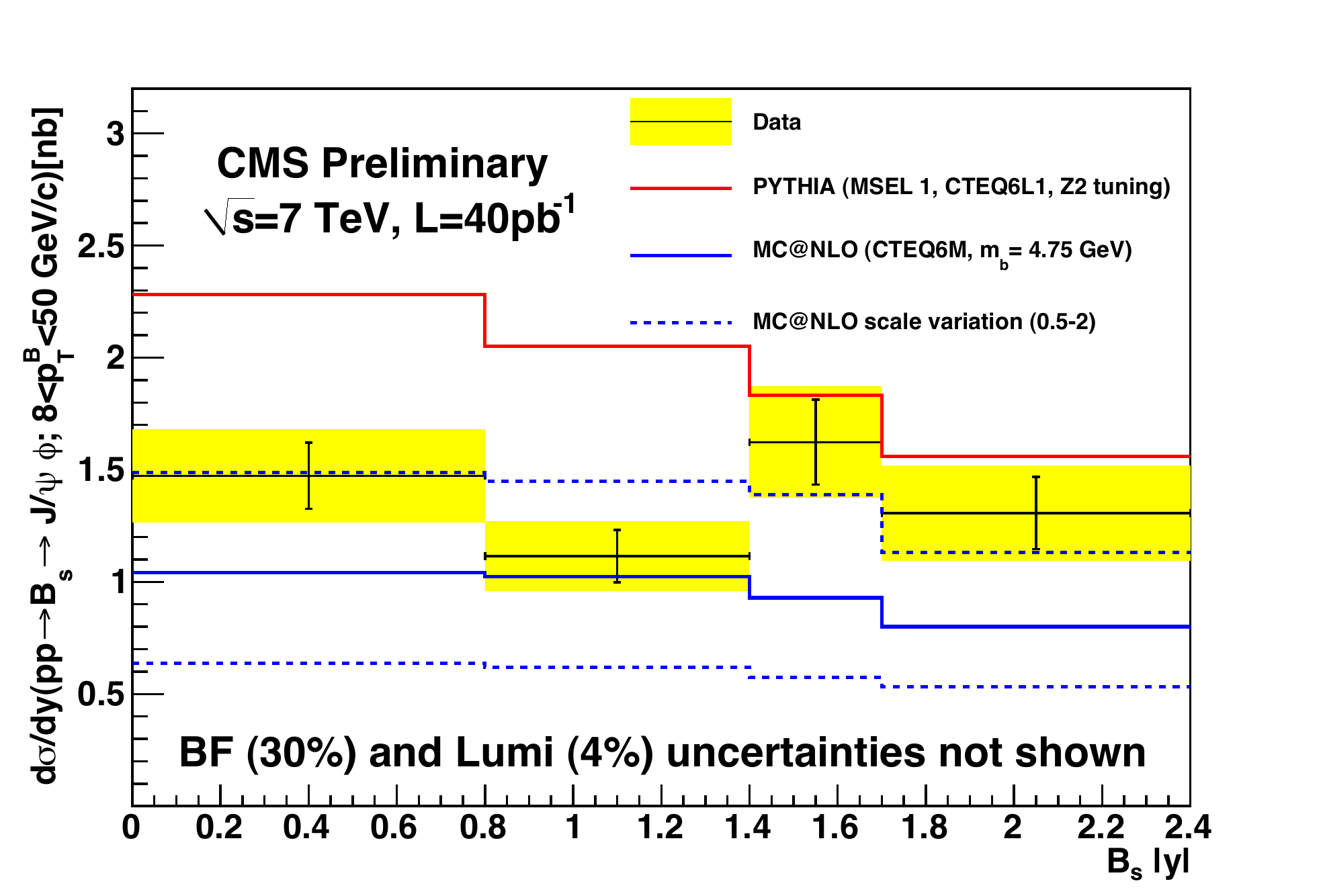}
\caption{
\small{Measured differential cross section $d\sigma /dp_\mathrm{T}$ 
(left--hand side) and $d\sigma /dy$ (to the right), 
for $B^+$ (upper row), $B^0$ (center) and $B_s$ (bottom) mesons, compared to the theoretical predictions.
 The error bars correspond to the statistical uncertainties, and the (yellow) band represents the uncorrelated systematic uncertainties. Overall uncertainties from the luminosity and  the branching fractions, reported in the plots, are not shown. The solid and dashed (blue) lines are the MC@NLO  prediction and its uncertainty, respectively. The dotted (red) line is the PYTHIA prediction.}
}
\label{fig:bHadrXsect}
\end{figure}

Combined fit to the B--meson masses and lifetimes are used to reject the background, 
with most of the fit shapes derived directly from data;
B--meson reconstruction efficiencies were taken from MC, whereas tracking,
muon and trigger efficiencies were derived from data.
The 2D maximum likelihood fits were performed in different bins of the mesons' \pt
and $y$, in order to measure the differential production cross sections.
The results of the measurements, shown in Fig.~\ref{fig:bHadrXsect}, are compared with the predictions of PYTHIA MC and MC@NLO calculation.
In general the agreement with the latter is fair, but some 
differences in the $y$ distributions are observed.

\subsection{Observation of other heavy hadrons}
\begin{figure}[htb!]
\centering
\includegraphics[width=0.32\textwidth,height=4.7cm]{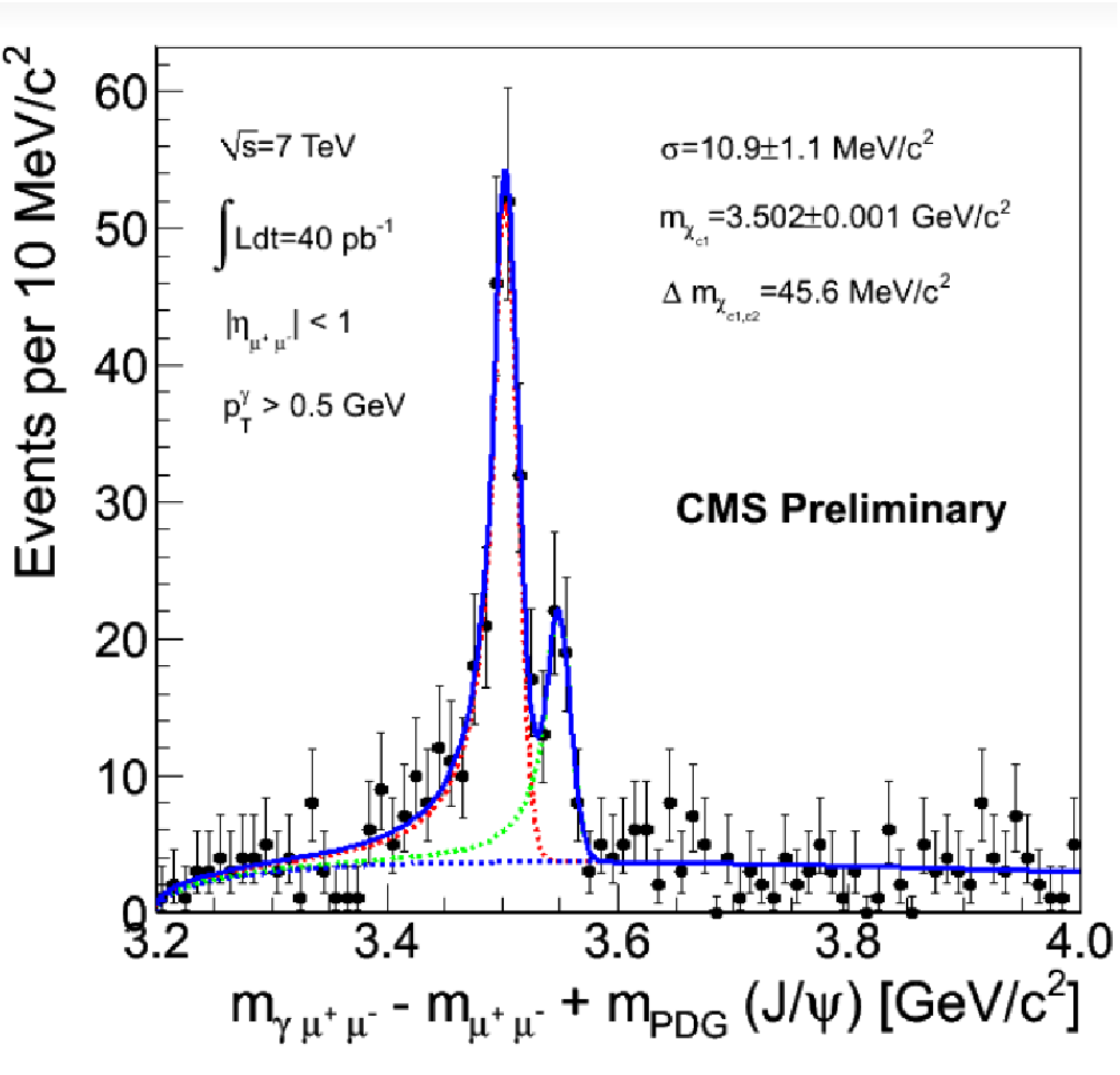}%
\includegraphics[width=0.32\textwidth,height=4.8cm]{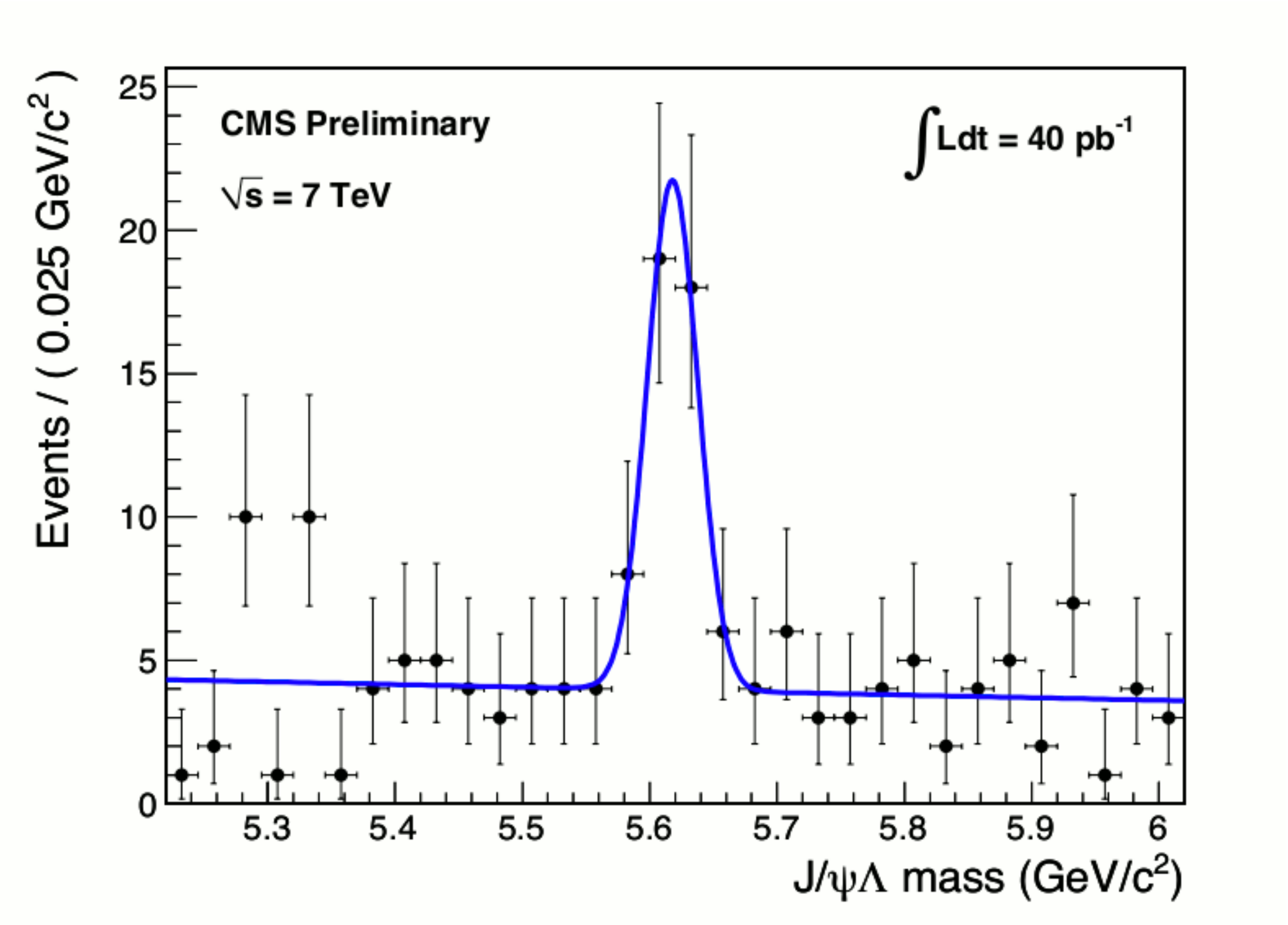}%
\includegraphics[width=0.32\textwidth,height=4.5cm]{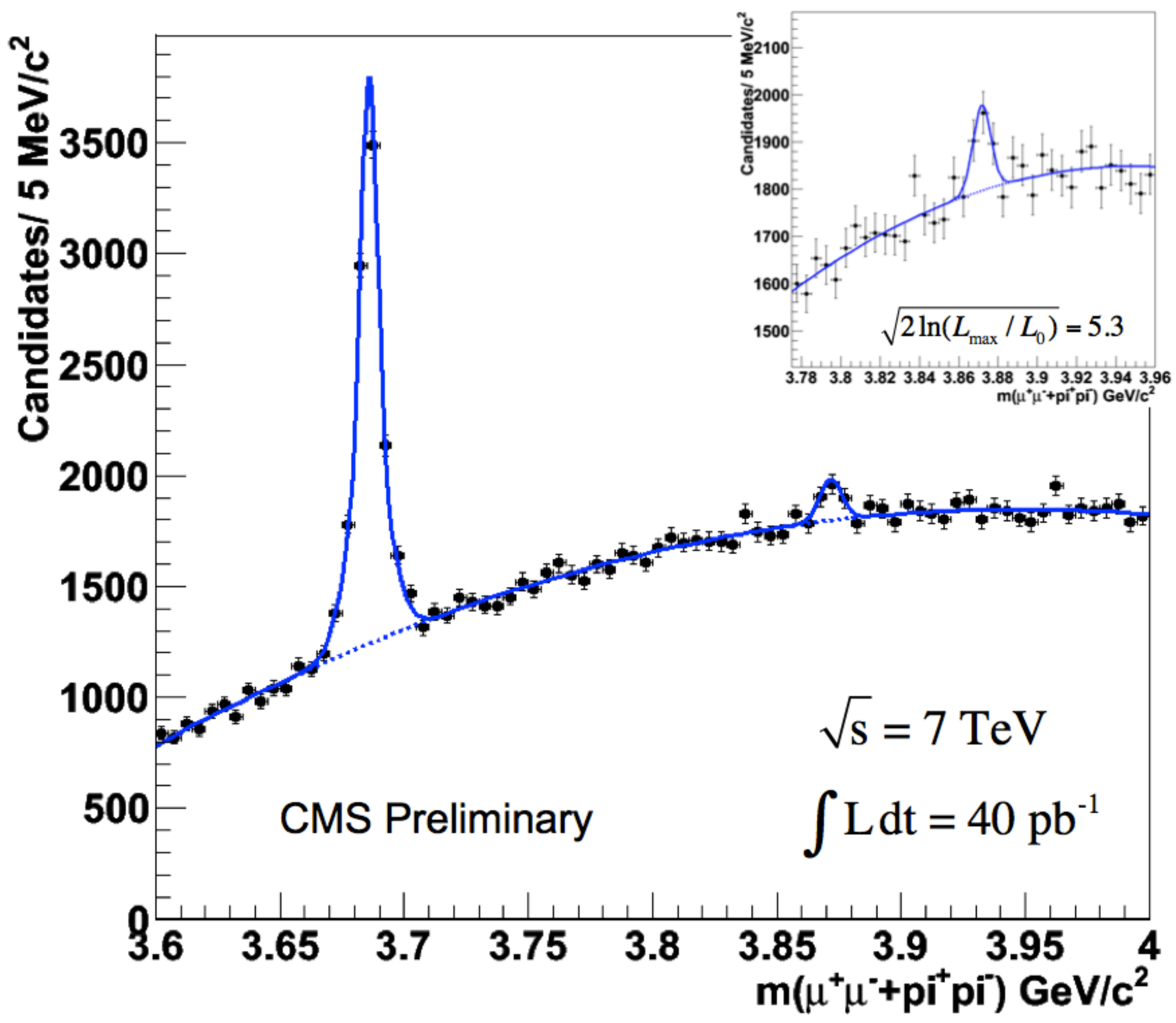}
\caption{
\small{Invariant mass peaks of the  $\chi_{c_1}$ and $\chi_{c_2}$
states (left), $\Lambda_b$ (center), the $\psi(2S)$ and the X(3872) states (right). The reconstructed decay channels are reported in the text.}
}
\label{fig:baryons}
\end{figure}

Beside B--mesons, other interesting particles have been observed and measured 
in the $40$ pb$^{-1}$ of $pp$ collision data collected in 2010 by the CMS detector.
The invariant mass peaks of the $\chi_{c_1}$ and $\chi_{c_2}$ states, separated by a $\Delta m \simeq 45$ MeV,
have been reconstructed through the decay channel $\chi_c \rightarrow J/\psi +\gamma $~\cite{cms:Chi_c}.
The $\Lambda_b$ baryon has also been observed through its decay to $J/\psi +\Lambda$~\cite{cms:Lambda_b}.
Furthemore, the puzzling X(3872) state has been observed in the decay channel to 
$J/\psi ~\pi^+ \pi^-$, and the ratio of its cross section over the $\psi(2S)$ has been measured,
giving: $R  = \textrm{X(3872)}/\psi(2S)= 0.087 \pm 0.017\textrm{(stat.)} \pm 0.009\textrm{(syst.)}$~\cite{cms:X_3872}. 
Fig.~\ref{fig:baryons} shows the invariant mass peaks of the aforementioned states.

\section{Open beauty}
b-­‐jets play a key role in searches
for new physics beyond the Standard Model.
It took a while to fully establish a consistency
between the Tevatron data and pQCD predictions for b--jet production cross section. 
Generally speaking, b--jets cross section measurements are highly non-trivial, and 
sizable uncertainties affect both theory and experiment: on one side, 
one has to deal with a typical multi--scale problem, in which 
the center--of--mass collision energy, mass of the b--quark and the factorization and re--normalization scales are entangled in a subtle way; 
on the other hand, excellent performance of the tracking is required, challenging the detector full potential. 

The CMS Collaboration has published results on b--jet production,
which are briefly reviewed in the next sections:
two complementary cross section measurements, making use of two different
b--tagging techniques, and a study of the
\BB correlations, performed over $85$ nb$^{-1}$, $60$ nb$^{-1}$, and $3.1$ pb$^{-1}$ of 2010 data, respectively~\cite{cms:bJets-pTrel,cms:bJets-SV,cms:BBcorrel}.
These results were obtained using track or particle--flow jets.
Typical values for the CMS performance at the time of the results presented here were:
 jet resolution about $10-15$\%, energy scale uncertainty below $3$\%.

\begin{figure}[htb!]
\centering
\includegraphics[angle=90, width=0.55\textwidth]{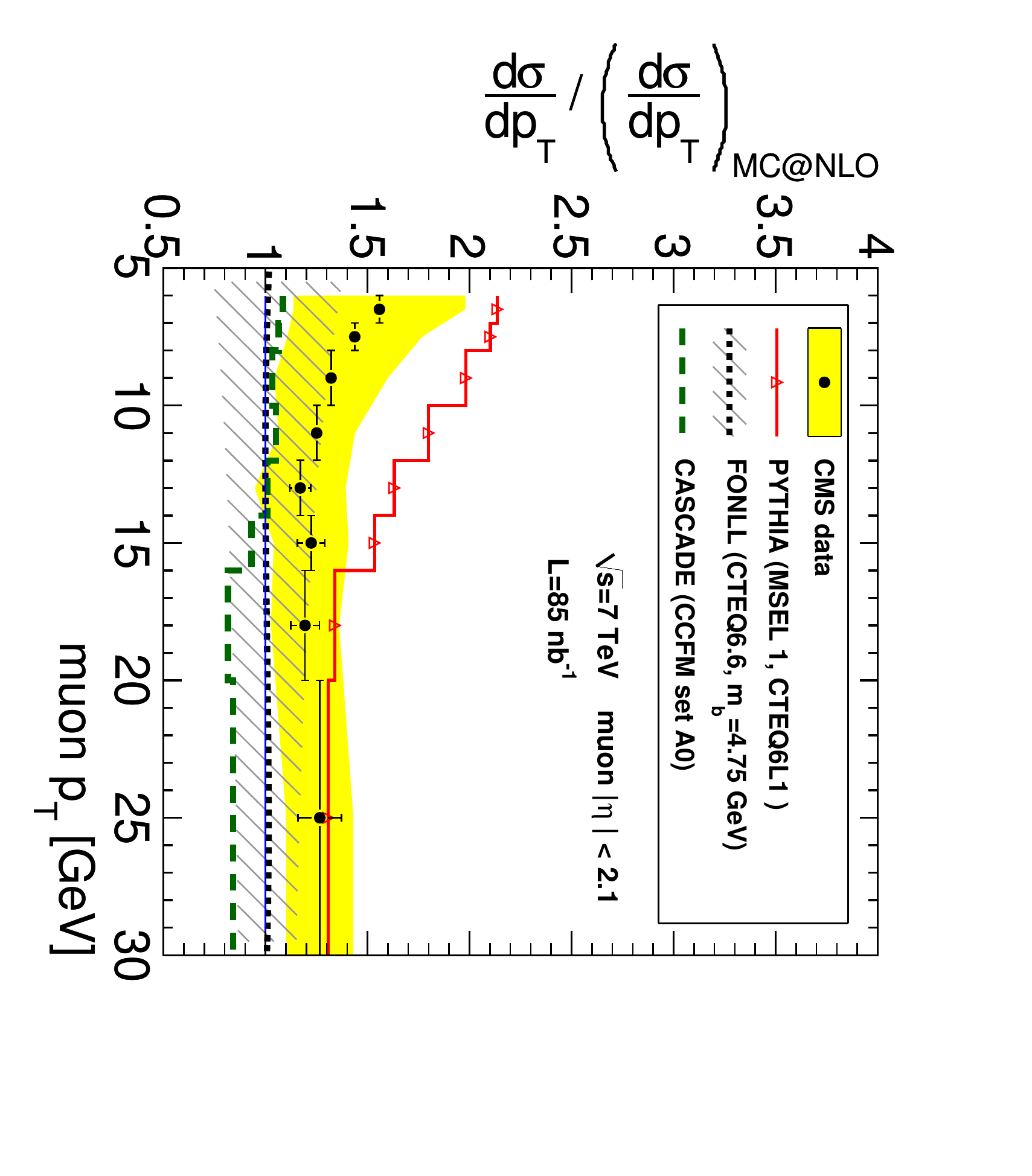}%
\includegraphics[angle=90, width=0.55\textwidth]{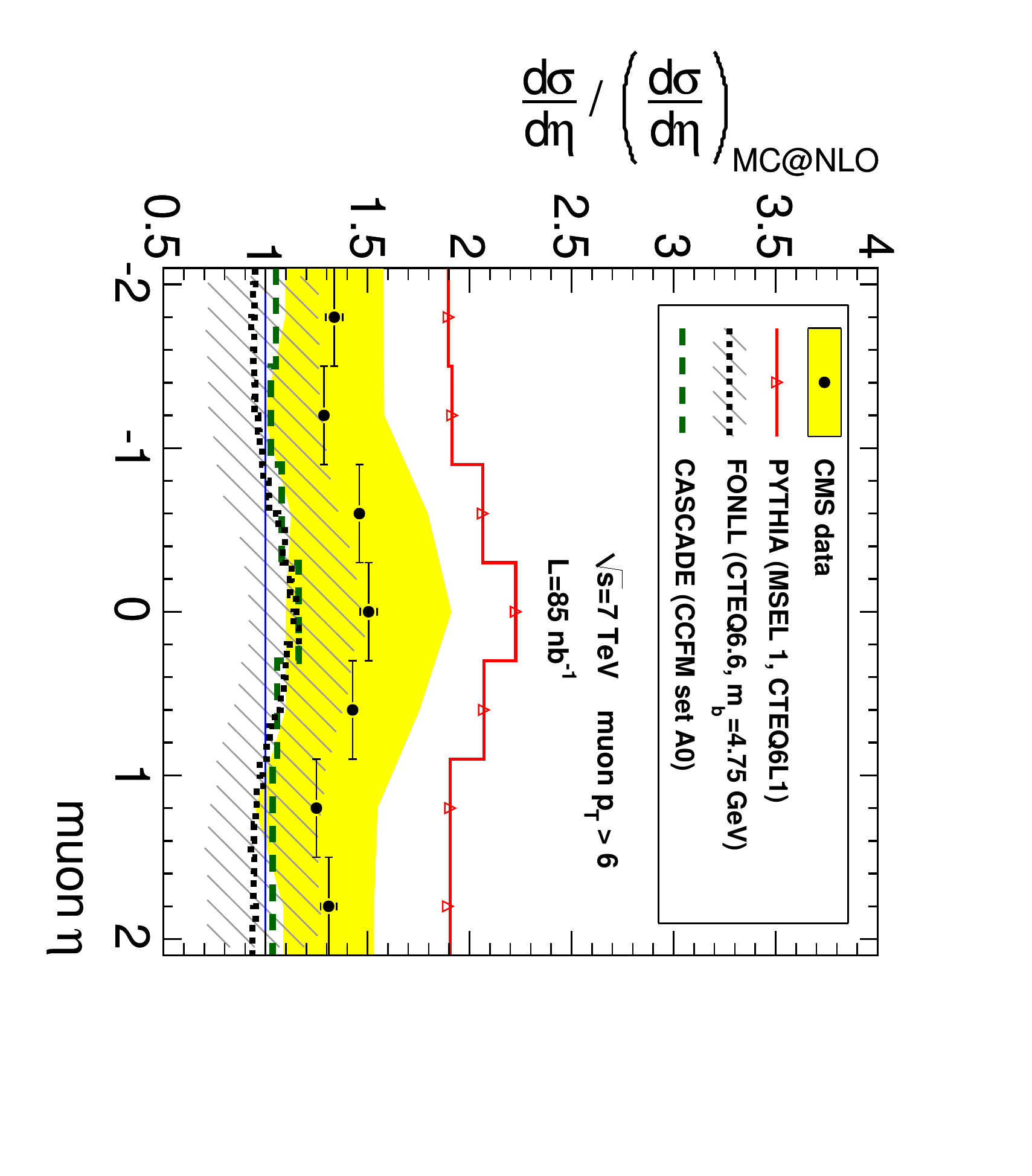}
\caption{
\small{Ratio of the measured production cross sections 
over the predictions of the MC@NLO computation, as a function of the muon's \pt
on the left hand side, and $\eta$ on the right, compared 
with the predictions provided by the Pythia, CASCADE MC 
and the FONLL calculations.}}
\label{fig:pTrel-ratios}
\end{figure}

\subsection{b--jets with muons}
The measurement of the integrated
and differential cross section of the reaction
$pp \rightarrow b+X \rightarrow \mu+X$, 
has been performed on jets coming from b--quarks.
The production of a b--quark decaying semi--leptonically is deduced by the identification of a rather energetic muon inside a jet, where the transverse momentum
relative to the jet axis is quite sizable.
For a muon from a b--decay, the transverse momentum
relative to the jet axis is on average larger 
than when the muon comes from light quarks; through this property it is hence 
possible to discriminate events in which b--quarks were produced. 

A binned log-likelihood fit is performed on the spectrum of such a quantity, called ``$p_{\mathrm{T}}^{rel}$'', 
using template distributions provided by the simulation for $b$ and $c$ quarks, and
derived from the data for gluons and light quarks.
This latter is dominated by hadrons misidentified as muons (mainly decay-in-flight), 
so they are reweighted by the misidentification rate measured in the data.
Considering that the fit is not able to distinguish light
quark, gluon and charm components, 
these are merged together. 

The fit stability was tested against variation of the binning, repeating the fits on many simulated pseudo--experiments, cross--checking the results using jets from particle flow and performing the fits on the impact parameter distribution. 

The b--jet tag efficiency achieved with this technique is about $74$\% at \pt$(\mu) \simeq 6$ GeV, and close to $100$\% above $20$ GeV, whereas the contamination is $\sim 7$\% in lowest \pt bin, asymptotically decreasing
towards $2$\% at high $p_{\mathrm{T}}$.
The uncertainties dominating the measurement are those coming from the approximate knowledge of the signal and the background $p_\mathrm{T}^{rel}$ shape.

Fig. \ref{fig:pTrel-ratios} shows the ratio of the measured differential production cross sections in \pt and $\eta$ over the predictions of the MC@NLO computation, as well as 
those provided by the Pythia, CASCADE MC and the FONLL calculations.

\begin{figure}[h!]
\centering
\includegraphics[width=0.49\textwidth]{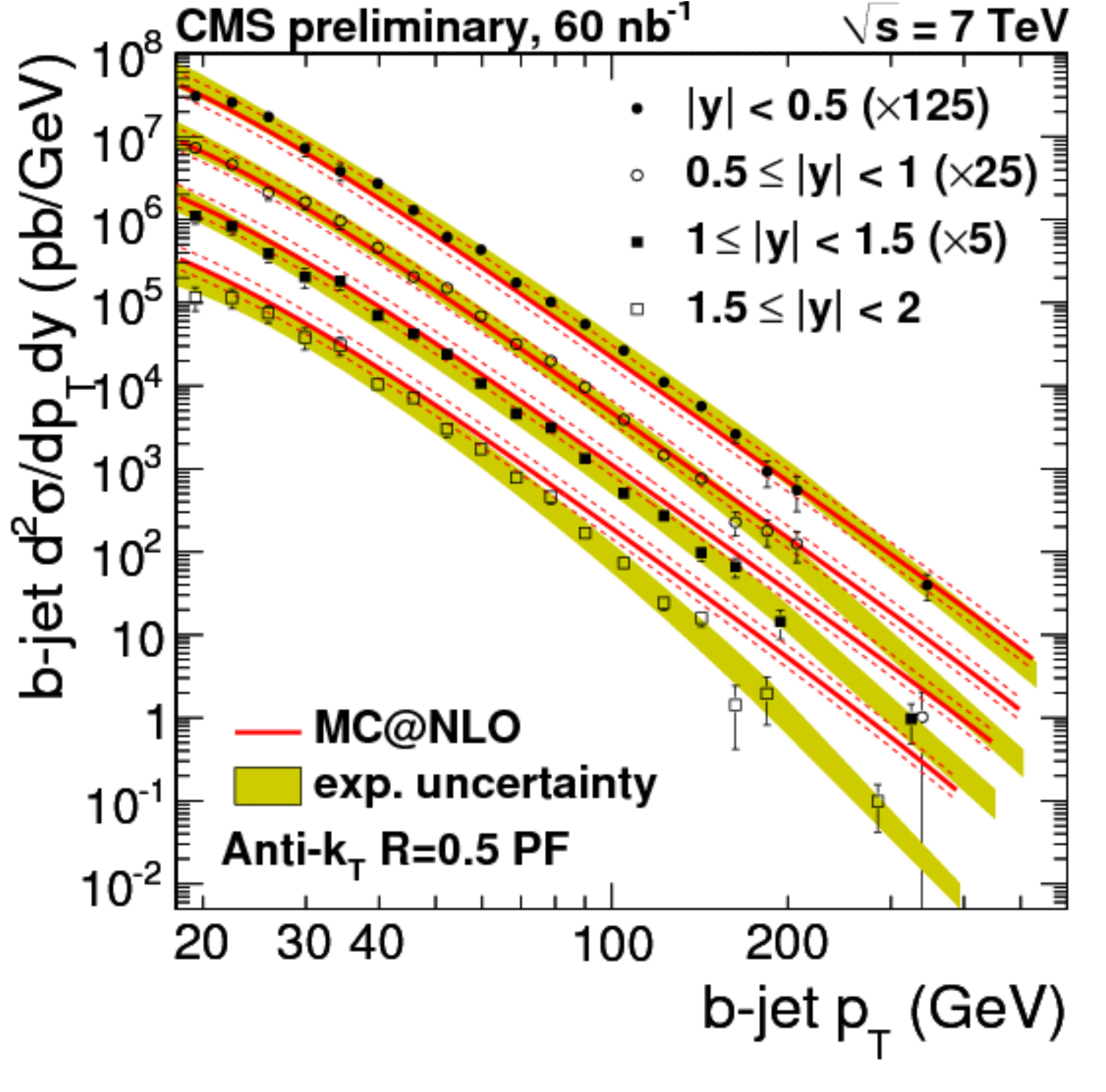}%
\includegraphics[width=0.49\textwidth]{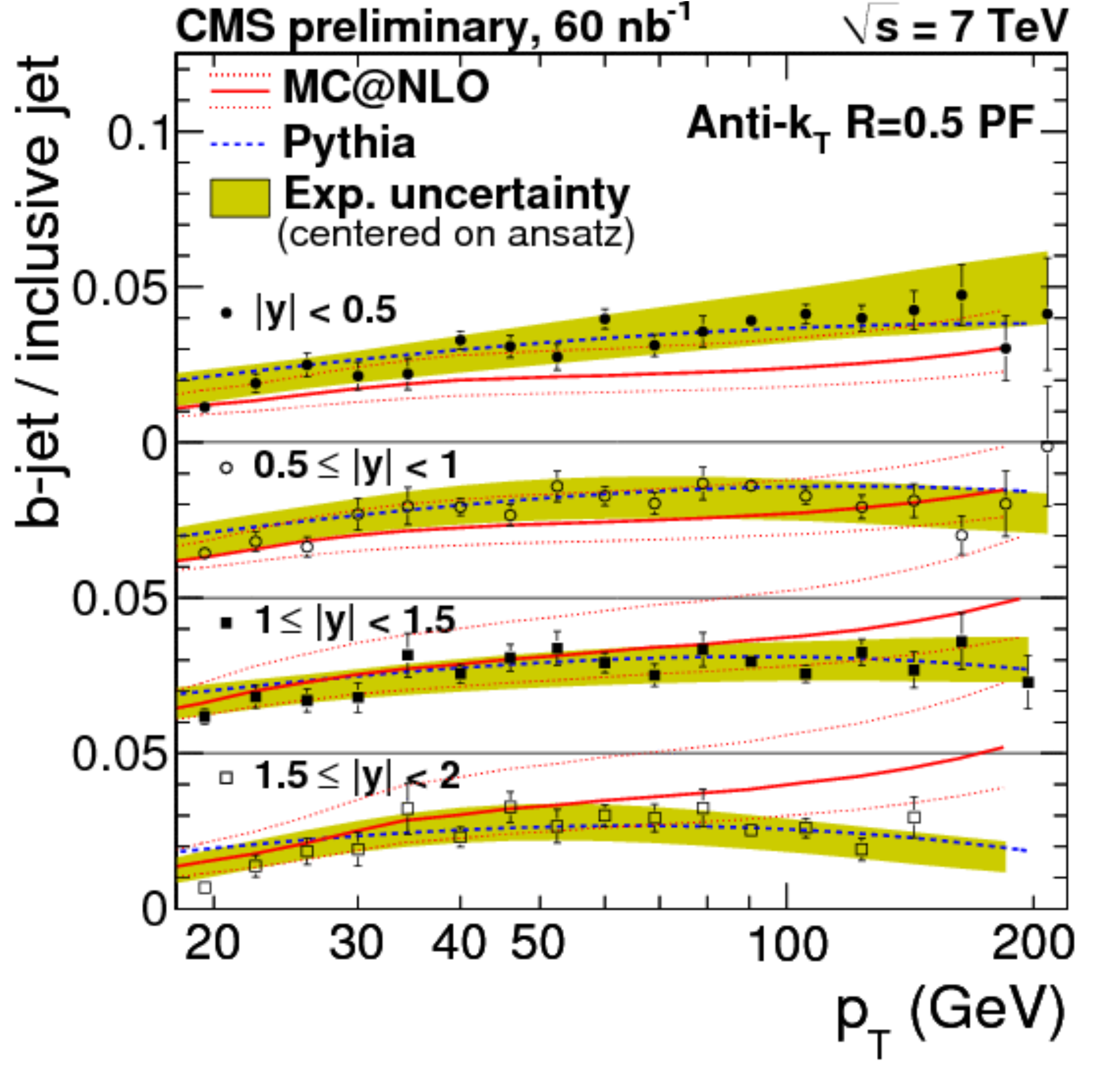}
\caption{
\small{Production cross section measurement for  b--jet as a function of the b--jet transverse momentum 
compared with the MC@NLO predictions (left) and the ratio of the b--jet cross section over the inclusive jets (right).}}
\label{fig:bjetXsec}
\end{figure}

\subsection{b--tagging with secondary vertices}
The identification of jets coming from the hadronization of b--quarks is possible 
also through the reconstruction of secondary vertices (SV).
Once a displaced SV is reconstructed, different discriminators 
can be used to tag the jet as originating 
from a b--quark. In the analysis presented here, the discriminator adopted is a 
monotonic function of the 3D decay length.
The decay length significance
cut is chosen so that the corresponding
tagging efficiency is about $60$\% at $p_\mathrm{T}^{jet} = 100$ GeV, 
with a  contamination of $\sim0.1$ \%.
The b--tagging efficiency and the mistag rates from $c$ or light jets are evaluated from simulated events
and constrained by a data/MC scale factor obtained from data.
The jet energy corrections for rapidity dependence, and those for absolute scale and \pt dependence,  come from real data and simulated events respectively.
In order to evaluate the purity of the selected sample, 
a fit to the SV mass distribution is performed, 
taking the shapes from simulated events, and letting free
the relative normalizations for c and b jets, with the (small) contribution from light quarks
fixed to the Monte Carlo expectations (``template fit''). \\

\begin{figure}[htb!]
\centering
\includegraphics[width=0.49\textwidth]{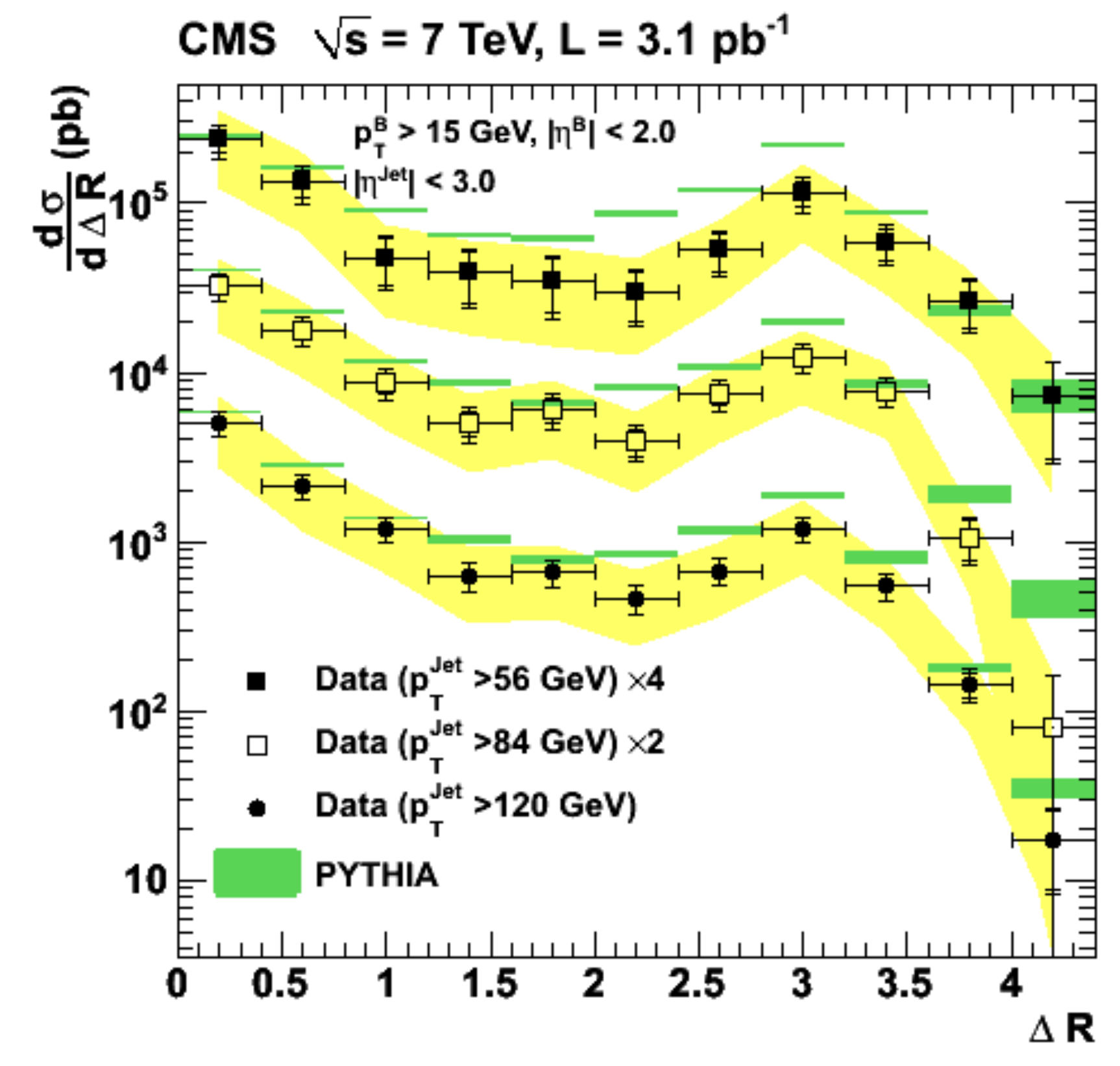}%
\includegraphics[width=0.49\textwidth]{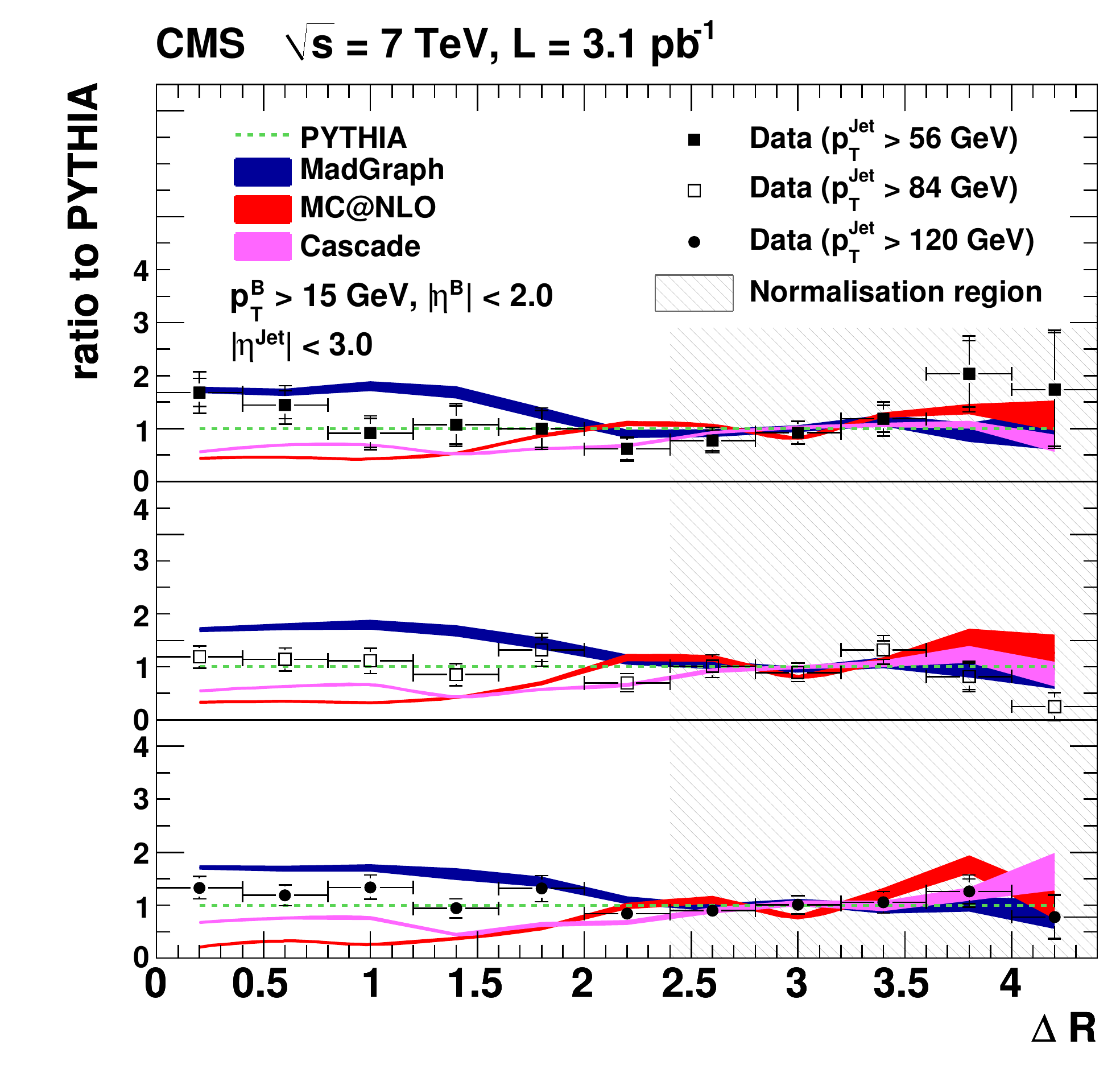}
\caption{
\small{Production cross sections as a function of the opening angle $\Delta R$ between the two B mesons
in different ranges of the leading jet $p_\mathrm{T}$, with absolute normalization (left);  
ratio of the measured cross section with the predictions from Pythia, compared with
MC@NLO, MadGraph and Cascade MC model predictions (right).
The dashed area at high $\Delta R$ values is used as the normalization region. }}
\label{fig:BBXsect}
\end{figure}

The data sample was collected using hadronic triggers with different thresholds of the jet  $p_\mathrm{T}$; to merge them, the individual \pt spectra of jets have been normalized with the luminosity of their data taking periods, and 
then combined into a single spectrum with the jet \pt bins corresponding to intervals
where the triggers were fully efficient.
 The overall transverse jet energy range goes from $18$ to 
$300$ GeV, and the measurements have been performed in four $\eta$ intervals. 

The leading systematic uncertainties are:
\vspace{-0.2cm}
\begin{itemize}
\item the jet energy scale of b--jets relatively to the inclusive ones ($4–5$\%);
\vspace{-0.2cm}
\item data-driven constraints on b-tagging efficiency ($20$\%);
\vspace{-0.2cm}
\item mistag rate for charm ($3–4$ \%) and for light jets ($1$ –- $10$\%).
\end{itemize}

Fig. \ref{fig:bjetXsec} shows the results for the production cross section measurements 
for b--jets as a function of the b--jet transverse momentum, compared 
with the MC@NLO predictions, and the ratio with the inclusive jets cross-section.
While the agreement with Pythia and MC@NLO is reasonable, significant differences in shape 
are evident, the simulations predicting more b--jets at high \pt than what is observed.

\begin{figure}[htb!]
\centering
\includegraphics[angle=90,width=0.49\textwidth]{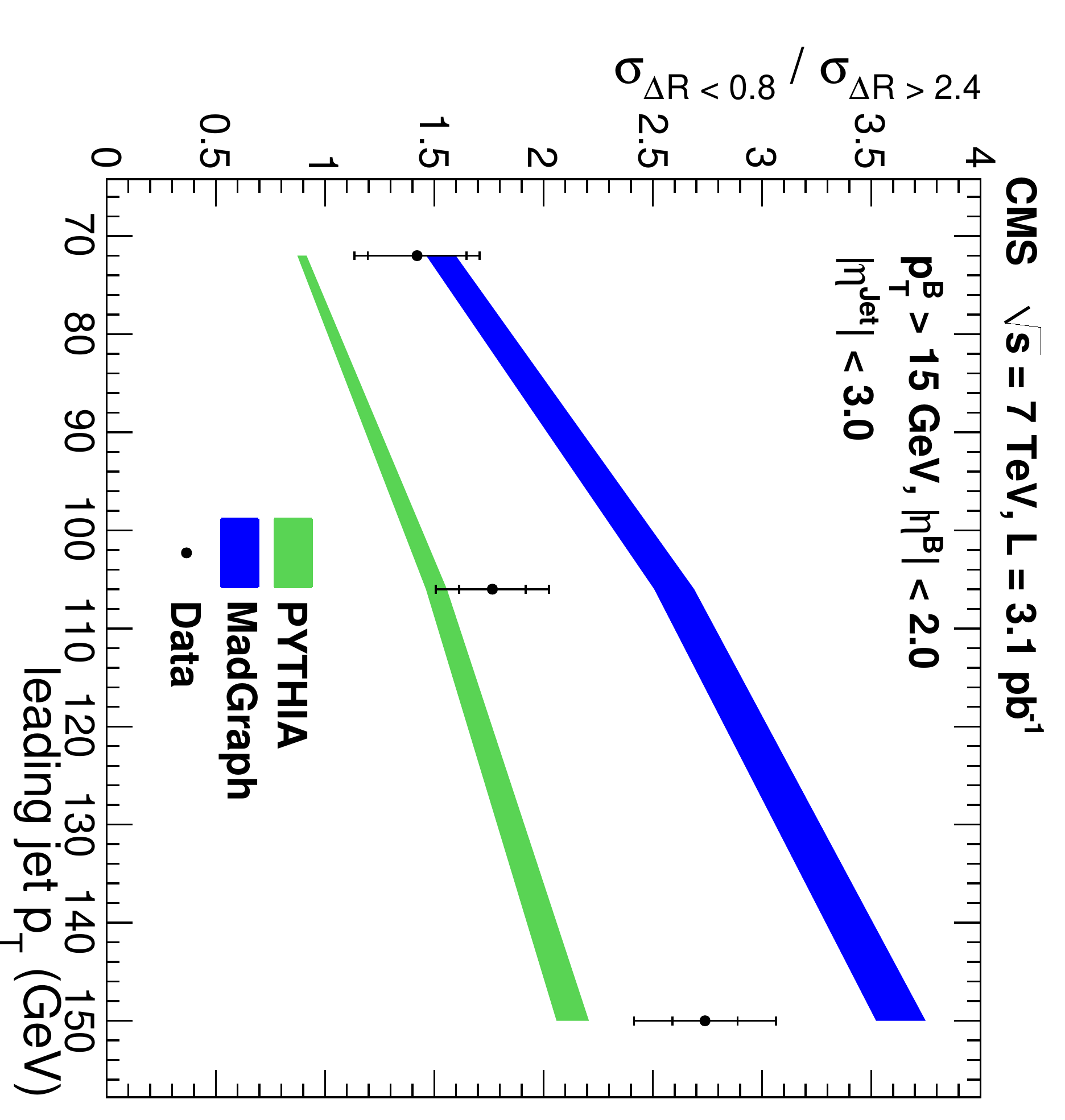}%
\includegraphics[angle=90,width=0.49\textwidth]{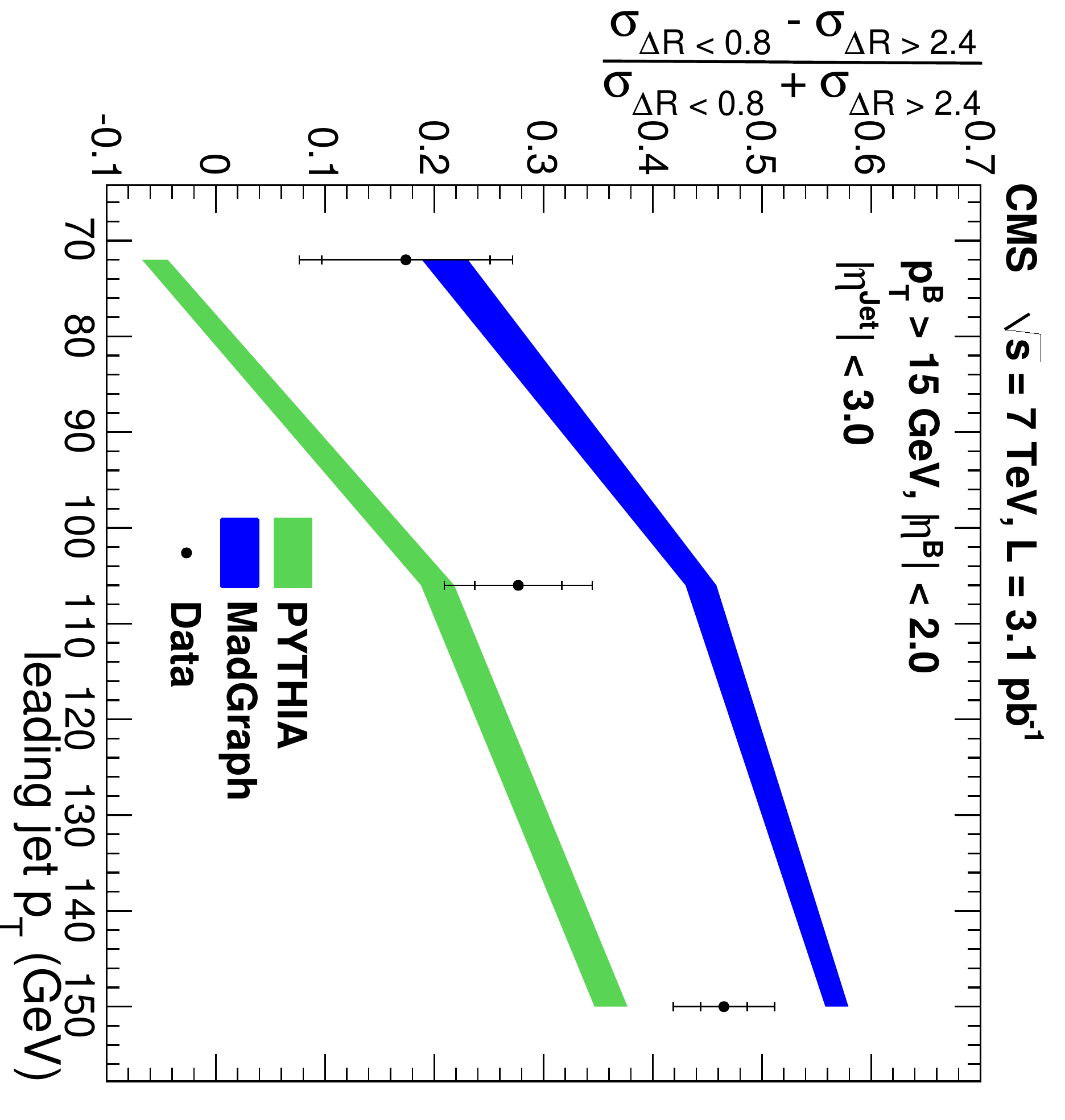}
\caption{
\small{Ratio between \BB production cross section in $\Delta R<0.8$ and $\Delta R> 2.4$ 
as function of leading jet $p_\mathrm{T}$, compared with Pythia, in green, and MadGraph MC predictions, in blue (left).
Asymmetry between \BB production cross section in $\Delta R<0.8$ and $\Delta R> 2.4$ 
as function of leading jet \pt(right).
}}
\label{fig:BBXsecRatioz}
\end{figure}

\subsection{\BB correlations}
Among the parton--level mechanisms of \bb production most relevant at LHC, there are the flavor creation,  of order $\alpha_s^2$, in which the heavy quarks are produced directly from two primary gluons 
in a t--channel process, and, at order $\alpha_s^3$, the so--called flavor excitation and the gluon splitting.
In the latter one, the \bb comes from a single gluon through an s--channel process. 

Considering B--hadrons coming from the hadronization of b--quarks, one important difference 
among the processes of direct flavor creation and the gluon splitting comes from the kinematic of the B's. The opening angle between the two B mesons is expected to be rather different, being larger in the process of lowest order. 

The CMS Collaboration has performed a detailed study of the \BB correlations,
providing a significant test of QCD and a further insight into the dynamics of \BB production mechanisms.
Making use of the SV--based b--tag techniques, the quantity 
$\Delta R$ defined as the distance in the $\eta$--$\phi$ plane between
the primary vertex and each SV, $\Delta R = \sqrt{\eta^2+\phi^2}$, is used to study the angular correlation of the B mesons. 

The production cross sections have been measured as a function 
of the opening angle $\Delta R$ among the two mesons; 
the results are presented in different ranges of the leading jet's $p_\mathrm{T}$, as in Fig. \ref{fig:BBXsect}, 
where also the ratio with the predictions from Pythia is presented, together with the comparison
with the MC@NLO, MadGraph and Cascade MC predictions. 
The dashed region at large $\Delta R$, in principle better known since it is
dominated by LO processes, is used in this case to obtain the normalization of
the remaining region of $\Delta R$. 

Considering that the flavor creation is expected to dominate at high values in
$\Delta R$, whereas the gluon splitting contributes more at low values,
it is interesting to check the agreement between the data and the theoretical models. 
This is shown in Fig. \ref{fig:BBXsecRatioz}, where the ratios of the cross sections in the regions $\Delta R < 0.8$ and  $\Delta R > 2.4$, and the relative asymmetry, are shown. 
None of the predictions describes the data very well, that lie between the MadGraph and Pythia. 
The calculations from MC@NLO do not describe the shape of the $\Delta R$ distribution either, in particular at small values, and the Cascade predictions are significantly below the data everywhere.

\section{Conclusions}
Various recent results published by the CMS Collaboration have been summarized, all 
of them obtained analyzing the $pp$ collision data at $7$ TeV 
collected in the year 2010. 

Measurements have been performed on quarkonia states, studying
the production of \jpsi and \upsi states, extracting the total and differential 
production cross sections as function of the mesons' \pt and $y$
for different polarization scenarios.  
The cross section of \jpsi coming from b--hadrons was also
measured and compared with the available theoretical computations.
The latest result on the \upsi polarization by the CDF Collaboration has also 
been presented. 
Exclusive reconstructions of different \mbox{B--mesons} such as
B$^+$, B$^0$ and B$_s$ have been presented, together with the production
cross section measurements versus \pt and $y$.
The observation of other beauty and charmed hadronic states, as the $\chi_{c_{1,2}}$, the $\Lambda_b$ baryon and the X(3872) state has also been reported. \\
Finally, the CMS results on open beauty production have been summarized, showing
the studies made with two different techniques for b--tagging, 
together with those on the angular correlation of \mbox{B--mesons}, 
and comparing the data with the predictions of the available theoretical models.

\vspace{1cm}

\newpage

\Acknowledgements
The involvement in this work was made possible by the founding
of the INFN, sezione di Padova.
I am grateful to Prof. U. Gasparini for all the sustaining and encouraging, and to 
the entire CMS Padova group for the fruitful collaboration.

\end{document}